\documentclass{aa}

\usepackage{graphicx}
\usepackage{siunitx}
\usepackage{comment}						 %
\includecomment{comment}
\specialcomment{note}{\begingroup\sffamily\color{gray}\footnotesize}{\endgroup}

\newcommand{\rev}[1]{#1}

\usepackage[disable]{todonotes}			%

\usepackage{txfonts}

\usepackage[colorlinks]{hyperref}
\hypersetup{colorlinks=true,linkcolor=blue,citecolor=blue,filecolor=blue,urlcolor=snsgreen} %

\DeclareSIUnit\mEarth{M_\oplus}
\DeclareSIUnit\mSun{M_\odot}
\DeclareSIUnit\rEarth{R_\oplus}
\DeclareSIUnit\year{yr}
\DeclareSIUnit\au{au}
\DeclareSIUnit\dex{dex}

\newcommand{\SE}{(super-)Earths}

\defcitealias{Emsenhuber2020}{Paper I}
\def\paperone{\citetalias{Emsenhuber2020}}
\defcitealias{Emsenhuber2020b}{Paper II}
\def\papertwo{\citetalias{Emsenhuber2020b}}
\defcitealias{Schlecker2021}{Paper III}

\defcitealias{Burn2021}{Paper IV}

\begin{document}

\title{The New Generation Planetary Population Synthesis (NGPPS)}
\subtitle{V. Predetermination of planet types in global core accretion models}
\author{M. Schlecker\inst{\ref{inst:mpia}}
  \and D. Pham\inst{\ref{inst:toronto},\ref{inst:sagan},}
  \and R. Burn\inst{\ref{inst:unibe},\ref{inst:mpia}}
  \and Y. Alibert\inst{\ref{inst:unibe}}
  \and C. Mordasini\inst{\ref{inst:unibe}}
  \and A. Emsenhuber\inst{\ref{inst:ualpl},\ref{inst:unibe}}
  \and H. Klahr\inst{\ref{inst:mpia}}
  \and Th. Henning\inst{\ref{inst:mpia}}
  \and L. Mishra\inst{\ref{inst:unibe},\ref{inst:unige}}
          }

   \institute{Max-Planck-Institut für Astronomie, Königstuhl 17, 69117 Heidelberg, Germany\\
        \email{schlecker@mpia.de} \label{inst:mpia}
      		\and
      	David A. Dunlap Department of Astronomy \& Astrophysics, University of Toronto, 50 St. George St., Toronto, ON M5S 3H4, Canada \label{inst:toronto}
        \and
   		Carl Sagan Institute, Cornell University, Ithaca, NY 14853, USA \label{inst:sagan}
   		\and
        Physikalisches Institut, University of Bern, Gesellschaftsstrasse 6, 3012 Bern, Switzerland \label{inst:unibe}
        \and
        Lunar and Planetary Laboratory, University of Arizona, 1629 E. University Blvd., Tucson, AZ 85721, USA \label{inst:ualpl}
        \and
	    Geneva Observatory, University of Geneva, Chemin Pegasi 51b, 1290 Versoix, Switzerland\label{inst:unige}        
    }

	\date{Received February 12, 2021; accepted April 20, 2021}

	\abstract
	{State-of-the-art planet formation models are now capable of accounting for the full spectrum of known planet types.
	This comes at the cost of increasing complexity of the models, which calls into question whether established links between their initial conditions and the calculated planetary observables are preserved.
	}
	{In this paper, we take a data-driven approach to investigate the relations between clusters of synthetic planets with similar properties and their formation history.}%
	{
	We trained a Gaussian Mixture Model on typical exoplanet observables computed by a global model of planet formation to identify clusters of similar planets. %
	We then traced back the formation histories of the planets associated with them and pinpointed their differences.
	Using cluster affiliation as labels, we trained a Random Forest classifier to predict planet species from properties of the originating protoplanetary disk.
	}
	{Without presupposing any planet types, we identified four distinct classes in our synthetic population.
	They roughly correspond to the observed populations of \mbox{(sub-)Neptunes}, giant planets, and \mbox{(super-)Earths}, plus an additional unobserved class we denote as ``icy cores''.
	These groups emerge already within the first \SI{0.1}{\mega\year} of the formation phase and are predicted from disk properties with an overall accuracy of $>\SI{90}{\percent}$.
	The most reliable predictors are the initial orbital distance of planetary nuclei and the total planetesimal mass available.
Giant planets form only in a particular region of this parameter space that is in agreement with purely analytical predictions.
Including N-body interactions between the planets decreases the predictability, especially for sub-Neptunes that frequently undergo giant collisions and turn into super-Earths.
	}
	{
	The processes covered by current core accretion models of planet formation are largely predictable and reproduce the known demographic features in the exoplanet population.
The impact of gravitational interactions highlights the need for N-body integrators for realistic predictions of systems of low-mass planets.
	}

	\keywords{Planets and satellites: formation -- protoplanetary disks -- Planets and satellites: dynamical evolution and stability -- Planet-disk interactions -- Methods: numerical -- Methods: statistical}

	\maketitle

	\section{Introduction}
One of the most remarkable findings in recent years of exoplanetology has been the enormous diversity of planetary systems~\citep[e.g.,][]{Ribas2007,Howard2012,Fressin2013,Petigura2013a,Mulders2015,Hobson2017a,Brewer2018,Owen2018b,Hsu2019,Bryan2019,He2020}. %
The rapidly increasing number of confirmed planets improves our ability to explore this diversity and to understand its origins.
To this end, a variety of physical mechanisms that influence the formation and evolution of planetary systems, and therefore shape their demographics, have been investigated.
Intensively studied mechanisms include the evolution of accretion disks~\citep[e.g.,][]{Lust1952,Lynden-Bell1974,Pringle1981}, their interaction with embedded planets that may result in orbital migration~\citep[e.g.,][]{Goldreich1979,Tanaka2002,DAngelo2003,Paardekooper2011,Dittkrist2014}, how these protoplanets form and grow by accreting solid components and gas~\citep[e.g.,][]{Bodenheimer1986,Ida1993,Pollack1996,Thommes2003,Fortier2013}, their gravitational interaction among each other~\citep[e.g.,][]{Chambers1996,Raymond2009}, photoevaporation of both protoplanetary disks~\citep{Hollenbach1994,Clarke2001,Alexander2014} and planetary atmospheres~\citep{Lammer2003,Owen2012b,Jin2014}, and the long-term evolution of planets and their atmospheres~\citep[e.g.,][]{Bodenheimer1986,Guillot2005,Fortney2010,Mordasini2012}.
While all these processes leave an imprint on the final planetary systems, observing them while they are in action has proven to be very challenging and was possible only in rare cases~\citep[e.g.,][]{Keppler2018}.
Global models of planet formation can mitigate this shortcoming by combining as many relevant physical processes as possible and simulating the growth and evolution of planets in an end-to-end fashion.
Thereby, they provide a link between properties of disks and observables of the resulting planets.
When employed within a Monte Carlo experiment with distributions of initial conditions, synthetic planet populations can be produced and statistically evaluated~\citep[e.g.,][]{Ida2004a,Mordasini2009a,Ndugu2017}.
Such population synthesis frameworks are increasingly able to produce different kinds of planets, from terrestrial-sized rocky planets to gas giants, using the same formation model.

The core accretion scenario~\citep{PerriCameron1974,Mizuno1978,Mizuno1980}, in which a solid planetary core forms that may subsequently accrete a gaseous envelope, has been recognized as the most common planetary formation avenue.
Concerning the problem of how this solid core grows, two different approaches have emerged:
commonly, the growth of the solid component has been modeled as the accretion of \mbox{$\sim$km-sized} planetesimals~\citep[e.g.,][]{Ida1993,Thommes2003}.
Under this assumption, the thresholds in the disk properties responsible for the emergence of different planet types are determined by the availability of planetesimals at the position of a growing planet and by the timescale for accreting them~\citep{Lissauer1987,Lissauer1993,Kokubo2000}.
In recent years, a growing body of literature includes the accretion of mm to cm-sized ``pebbles'', whose motions are decoupled from the gas disk~\citep{Ormel2010,Lambrechts2012,Bitsch2017}.
Here, the resulting radial motion of the particles causes an interrelation between the inner and outer regions of the disk~\citep{Morbidelli2012,Lambrechts2014a,Ormel2017}.

Both approaches have allowed the unambiguous predetermination of planetary parameters from initial conditions~\citep[e.g.,][]{Kokubo2002,Ida2004,Lin2018}.
However, with ever more sophisticated models of increasing complexity, it is uncertain whether these relationships persist.
In particular, the inclusion of N-body treatment of protoplanets could destroy these connections due to the chaotic component it introduces.
A number of studies have addressed this problem in different ways, either by categorizing the outcomes of simulations with different initial conditions~\citep{Mordasini2009a,Mordasini2012d,Bitsch2015b,Bitsch2018d,Miguel2019}, or by relating synthetic populations to the observed sample of exoplanets~\citep{Mordasini2009,Chambers2018,Fernandes2019,Mulders2020} or transitional disks~\citep{ChaparroMolano2019}.
A main limitation of these advances has been their restriction to a particular region of the planetary parameter space. %

Recent advancements of our formation model~\citep{Emsenhuber2020} now allow for an extension of these investigations to the full range of currently known planet types.
Therefore, in this study, we statistically assess the relations between a number of relevant disk properties and the emerging planet types in the context of the core accretion paradigm.
To this end, we investigate synthetic planet populations computed with the Generation III \textit{Bern} Model of planet formation and evolution~\citep[][hereafter Paper I]{Emsenhuber2020}.
Previous papers in this series have presented populations from this model with different numbers of planets per system~\citep[][Paper II]{Emsenhuber2020b} and varying host star masses (Burn et al., subm., Paper IV).%
Here, we focus on two populations of systems around solar-type stars: \textit{NG73} for isolated single planets, and \textit{NG76} with 100~planetary embryos growing concurrently~(\papertwo).
We thereby take care to follow a purely data-driven approach and do not presuppose planet types motivated by observations or theoretical arguments.

This paper is divided into six sections.
In Sect.~\ref{sec:formation_model}, we describe the formation model and introduce the synthetic planet populations.
We then present a cluster analysis performed on these populations in Sect.~\ref{sec:cluster_analysis}.
Section~\ref{sec:prediction} investigates to what degree the identified clusters of similar planets can be predicted from properties of protoplanetary disks.
In Sect.~\ref{sec:discussion}, we interpret our results and discuss their implications for planet formation.
We conclude by summarizing our findings in Sect.~\ref{sec:conclusion}.

\section{Planet population synthesis}\label{sec:formation_model}
This work analyzes synthetic planet populations for solar-mass host stars from the Generation~III Bern global model of planet formation and evolution~(\paperone).
The formation part of the model combines the evolution of a protoplanetary disk with both gas and solids components, the growth and determination of the internal structure of protoplanets, their dynamical interactions and gas-driven planetary migration.

The gas disk is modeled as a viscously accreting disk~\citep{Lust1952,Lynden-Bell1974,Pringle1981} with an $\alpha$-parametrization \citep{Shakura1973} for the turbulent viscosity.
The vertical structure is computed following \citet{Nakamoto1994} and \citet{Hueso2005} under an evolving luminosity of the star \citep{Baraffe2015}.
The solid disk component is modeled in a fluid-like description where the dynamical state of planetesimals is given by the stirring due to other planetesimals and protoplanets~\citep{Thommes2003,Chambers2006,Fortier2013}.

The formation of protoplanets follows the core accretion paradigm \citep{PerriCameron1974,Mizuno1978,Mizuno1980} with planetesimal accretion in the oligarchic regime \citep{IdaMakino1993}.
We calculate the structure of the planetary envelopes by directly solving one-dimensional internal structure equations~\citep{Bodenheimer1986}.
Initially, gas accretion is limited by the ability of the planet to radiate away the gravitational energy release by accretion of solids and gas \citep{Pollack1996,Lee2015}.
At this stage, the internal structure is used to compute the gas accretion rate.
Once a planet exhausts the supply from the gas disk (either because cooling becomes efficient or because the disk disperses), the envelope is no longer in equilibrium with the disk and contracts \citep{Bodenheimer2000}.
In this \textit{detached} phase, the internal structure equations are used to determine the planet's radius.
The formation stage also includes gas-driven planetary migration in the Type~I \citep{Paardekooper2011} and Type~II \citep{Dittkrist2014} regimes.

The planetary seeds start with a mass of \SI{0.01}{\mEarth} and are inserted with random initial orbital distances $a_\mathrm{start}$ drawn from a log-uniform distribution between the inner disk edge and \SI{40}{\au}.
When multiple embryos are present in the same disk, their gravitational interactions are modeled during the first \SI{20}{\mega\year} using the \texttt{Mercury} N-body integrator~\citep{Chambers1999}.
After this time, the model switches to the evolutionary stage.
Here, the thermodynamical evolution is calculated for each planet individually up to a simulation time of \SI{10}{\giga\year}.
This stage includes atmospheric loss via photoevaporation~\citep{Jin2014} and tidal migration.
As a result, the model is able to compute the planets' masses, radii, and luminosities as a function of time.

For a thorough description of the Generation~III Bern Model and an outline of recent advancements of the framework~\citep{Alibert2005,Mordasini2009a,Mordasini2012a,Mordasini2012,Alibert2013}, we refer to~\paperone.

Synthetic planet populations are produced by running the model in a Monte Carlo scheme, where initial conditions are drawn randomly from distributions motivated by observational \citep{Santos2003,Lodders2003,Andrews2010,Venuti2017,Ansdell2018,Tychoniec2018} or theoretical constraints \citep{Drazkowska2016,Lenz2019}.
The distributed variables include the initial gas disk mass $M_\mathrm{gas}$, the inner edge of the disk $r_\mathrm{in}$, its dust-to-gas ratio $\zeta_\mathrm{d,g}$, the mass loss rate due to photoevaporative winds $\dot{M}_\mathrm{wind}$, and the starting locations of the planetary seeds $a_\mathrm{start}$.
The values or distributions of all model parameters are listed in Tab.~\ref{tab:modelParams} and are motivated in detail in \paperone\ and \papertwo.
\begin{table*}
        \caption{Choice of model parameters}             %
        \label{tab:modelParams}
        \centering                          %
        \begin{tabular}{r c c c}        %
                \hline\hline                 %
                Parameter & Symbol & Distribution & Range or Median$\substack{+84\% \\ -16\%}$\\
                \hline                        %
                \textbf{Fixed Parameters}& & & \\
                Stellar Mass & & -- & \SI{1}{M_\odot}\\
                Disk Viscosity & $\alpha$ & -- & $\SI{2e-3}{}$ \\
                Power Law Index (Gas) & $\beta_\mathrm{g}$ & -- & $0.9$\\
                Power Law Index (Solids) & $\beta_\mathrm{s}$ & -- & $1.5$\\

                Radius of Planetesimals & & -- & \SI{300}{\meter} \\
                Number of Planet Seeds & & -- & $1$ (\textit{NG73})/$100$ (\textit{NG76}) \\
                Mass of Planet Seeds & & -- & \SI{0.01}{M_\oplus} \\
                \hline
                \textbf{Monte Carlo Parameters}& & & \\
                Initial Gas Surface Density at \SI{5.2}{au} & $\Sigma_0$& log-normal & $132 \substack{+37 \\ -27}\,\SI{}{\gram\per\centi\meter\squared}$\\
				Dust-to-gas Ratio  & $\zeta_\mathrm{d,g}$ & log-normal & $0.02 \substack{+0.01 \\ -0.01}$\\
                Inner Disk Radius & $R_\mathrm{in}$ & log-normal & $ 4.74 \substack{+4.94\\ -2.42} \,\SI{}{\day}$ \\
                Gas Disk Cutoff Radius & $R_\mathrm{cut,g}$& log-normal & $ 56 \substack{+36 \\ -21}\,\SI{}{au}$\\
                Solid Disk Cutoff Radius & $R_\mathrm{cut,s}$& log-normal & $R_\mathrm{cut,g}/2$ \\
                Photoevaporation Efficiency & $\dot{M}_\mathrm{wind}$ & log-normal & $ (1.0 \substack{+2.2 \\ -0.7})\times 10^{-6}\,\SI{}{M_\odot/yr}$ \\
                Starting Position of Planet Seeds& $a_\mathrm{start}$ & uniform in $\log a$ & $R_\mathrm{in} \,\mathrm{to} \,40\,\mathrm{au}$\\
                \hline
                \textbf{Derived Parameters}& & & \\
				Host Star Metallicity& [Fe/H]& normal & $-0.03 \pm 0.20$\\
                Initial Gas Disk Mass & $M_\mathrm{gas}$ &  log-normal & $0.03 \substack{+0.04 \\ -0.02}$ \SI{}{M_\odot}\\
                Initial Solid Disk Mass& $M_\mathrm{solid}$ & $\sim$ log-normal & $95 \substack{+147 \\ -55}$\SI{}{M_\oplus}\\
                Disk Dispersal Time& $t_\mathrm{disk}$& -- & $(3.2 \substack{+1.9 \\ -1.0})\times 10^{6}\,\SI{}{yr}$ \\
                \hline
        \end{tabular}
        \tablefoot{Upper panel: parameters that are fixed for each simulation. Middle panel: distributions of Monte Carlo parameters that are drawn randomly. Lower panel: Quantities that are derived from or controlled by other parameters. The upper and lower limits denote 84th and 16th percentiles, respectively.}
\end{table*}

Our goal is to uncover characteristic links between these properties and the emerging planet types, which requires to robustly define the latter first.
This step may be impaired by the stochasticity of an N-body treatment that smears the boundaries between clusters of similar planets. 
We thus \rev{examine both a population with a single planet per system and a population with multiple planets per system. %
For the single-planet population}, called \textit{NG73}, 30,000 systems were simulated.
In 29,455 systems the planet was not accreted onto the star and is still present after \SI{5}{\giga\year}, which we consider as time of observation.

\rev{To consider the impact of gravitational interactions among planets, we investigate the multi-planet population~\textit{NG76} and compare it to the single-planet case.}
\rev{In each of its systems, an initial set of 100 protoplanets competed for material and interacted gravitationally.
	All other boundary conditions were left the same, and the Monte Carlo parameters were drawn from the same distributions.
	The N-body module integrated for \SI{20}{\mega\year} to cover the entire formation phase with planets still embedded in the disk, as well as an appropriate subsequent evolutionary era without disk interactions (\paperone).}
\rev{Out of the 1000 simulated systems, 32,030 planets survived until $t = \SI{5}{\giga\year}$.}
\rev{For detailed descriptions of both planet populations, see \papertwo{} and \citet{Schlecker2021}.}

\section{Cluster analysis}\label{sec:cluster_analysis}
A cluster analysis aims at identifying groups of entities that share similar properties in a specific set of parameters.
In our case, we aim to explore which distinct planet species emerge from our planet formation model and how they compare to observed \mbox{(exo-)planet} types.
Accordingly, we chose as training features three parameters typically obtained from exoplanet observations: the orbital semi-major axis $a$, the planet mass $M_\mathrm{P}$, and the planet radius $R_\mathrm{P}$.
Our clustering was done in a purely data-driven fashion and without any prior knowledge on existing or expected planet types.
The only information our clustering model received was a snapshot of our synthetic planet population at a simulation time of \SI{5}{\giga\year}.

\subsection{Data preparation}
In general, clustering methods are not scale-invariant~\citep{Jain1988}.
The application of cluster algorithms to unevenly scaled data~sets can thus lead to compromised results.
Based on the distribution of the parameters of interest in our data~set, we rescaled the features $a$, $M_\mathrm{P}$, and $R_\mathrm{P}$ by applying a $\log_{10}$.

\subsection{\rev{Model selection} and hyperparameters}\label{sec:clusteralgos}
\rev{We performed the clustering using Gaussian Mixture Models~\citep[GMM,][]{McLachlan1988}, a class of hierarchical, probabilistic clustering algorithms.
A GMM consists of multiple components $i=1\dotsb N$ of multivariate normal distributions, each characterized by its weight $\phi_i$, its mean $\mu_i$, and its covariance matrix $\Sigma_i$.
The model then takes the form
	\begin{equation}
		\sum_{i=1}^{N} \phi_i \mathcal{N}(\mu_i, \Sigma_i).
	\end{equation}
	During training on a data~set, the parameters $\phi_i$, $\mu_i$, and $\Sigma_i$ are updated using the expectation-maximization (EM) algorithm~\citep{Hartley1958}.
	A free hyperparameter is the number of Gaussian components $N$, that is, the number of Gaussian distributions the data points are assumed to be generated from.
	The trained GMM gives each data point a set of $N$ probabilities, corresponding to the probability that the data point belongs to a specific component $i$.
	When we classified our data, we assigned each planet the component (i.e., the planet cluster) with the highest probability.

Since GMM, and clustering algorithms in general, are unsupervised methods, the selection of a ``best'' model has to be seen in the context of the goal we want to achieve.
We aimed at identifying groups of planets based on overdensities in the planetary parameter space, regardless of their shape.
With this goal in mind, we have explored several other algorithms in addition to GMM and found that they consistently performed worse on our data~set~(see Appendix~\ref{sec:appendix_cluster_analysis}).
}
Using \rev{the \texttt{scikit-learn}~\citep{scikit-learn}} implementation \rev{of GMM} with default arguments, the only free hyperparameter was the number of \rev{clusters in the data $N$}.
\rev{In finding the optimal choice of $N$, we were aided by several validation metrics.
We considered the Akaike Information Criterion~\citep[AIC,][]{Akaike1973, Cavanaugh2019}, the Bayesian Information Criterion~\citep[BIC,][]{Schwarz1978}, the Davies-Bouldin score~\citep[DB,][]{Davies1979}, the Caliński-Harabasz score~\citep[CH,][]{Calinski1974}, and the Silhouette score~\citep{Rousseeuw1987}.
These metrics assess clustering performance with different approaches, and due to the complex structures in our data they can contradict each other.
We provide a detailed description of the different metrics in Appendix~\ref{sec:appendix_metrics}.
For now, it is important to note that AIC, BIC, and DB should be minimized, and CH and the Silhouette score should be maximized.}
	\begin{figure*}
		\centering
		\includegraphics[width=\hsize]{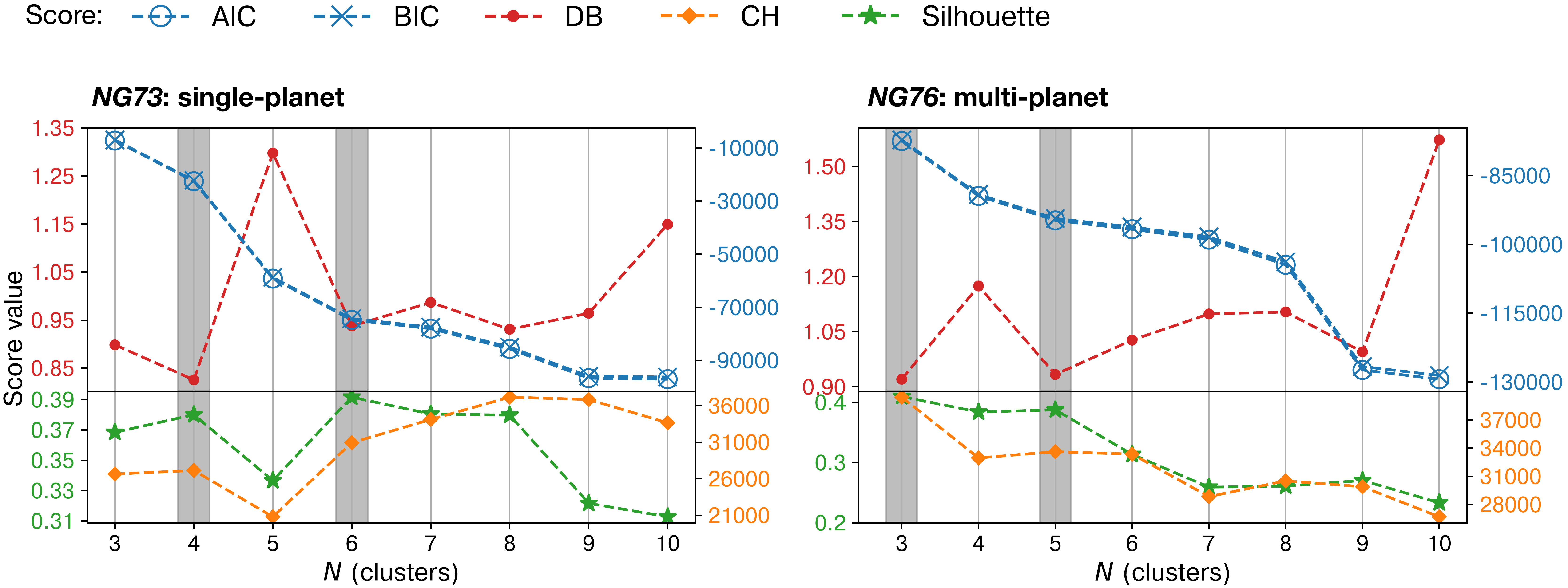}
		\caption{Validation scores for Gaussian Mixture Models \rev{with different numbers of components $N$.
		For AIC, BIC, and DB (top panels), lower values are preferred; and for Silhouette score and CH (bottom panels), higher values are preferred.}
		AIC and BIC \rev{generally} show indistinguishable values.
		Based on these scores, \rev{sensible choices are $N=4$ and $N=6$ for \textit{NG73}, and $N=3$ and $N=5$ for \textit{NG76} (highlighted in gray). Note the different y-axis scales.}
		}
		\label{fig:GMM_scores}
	\end{figure*}
In Fig.~\ref{fig:GMM_scores}, \rev{we show the different scores for GMMs with $N \in [3...10]$ upon applying them to our single-planet (\textit{NG73}) and multi-planet (\textit{NG76}) population, respectively.}
	\rev{For \textit{NG73},} two potential choices stick out, $N = 4$ and $N = 6$.
\rev{To decide between these options, we produced diagnostic scatter plots where all possible 2D~projections of the planetary parameter space are shown with planets color-coded by cluster affiliation.
The plots for the candidate models are shown in Fig.~\ref{fig:diagnostics_GMM}.
}
While human bias might be an issue at this step, we took care to judge the clustering only based on over- and underdensities of planets and not based on where we expected different planet types.
\rev{We found that the GMM with $N = 4$ performed best.}
\rev{For \textit{NG76}, both $N=3$ and $N=5$ yielded promising scores.
By judging the corresponding diagnostic plots, we concluded that $N = 5$ clusters is the preferred mode.
	}
\rev{With all hyperparameters fixed, we performed the unsupervised training of our nominal GMMs on the full data~sets and considering full covariance matrices.}

\subsection{\rev{Detected planet clusters}}\label{sec:results:planetclusters}
\begin{figure*}
	\centering
	\includegraphics[width=\hsize]{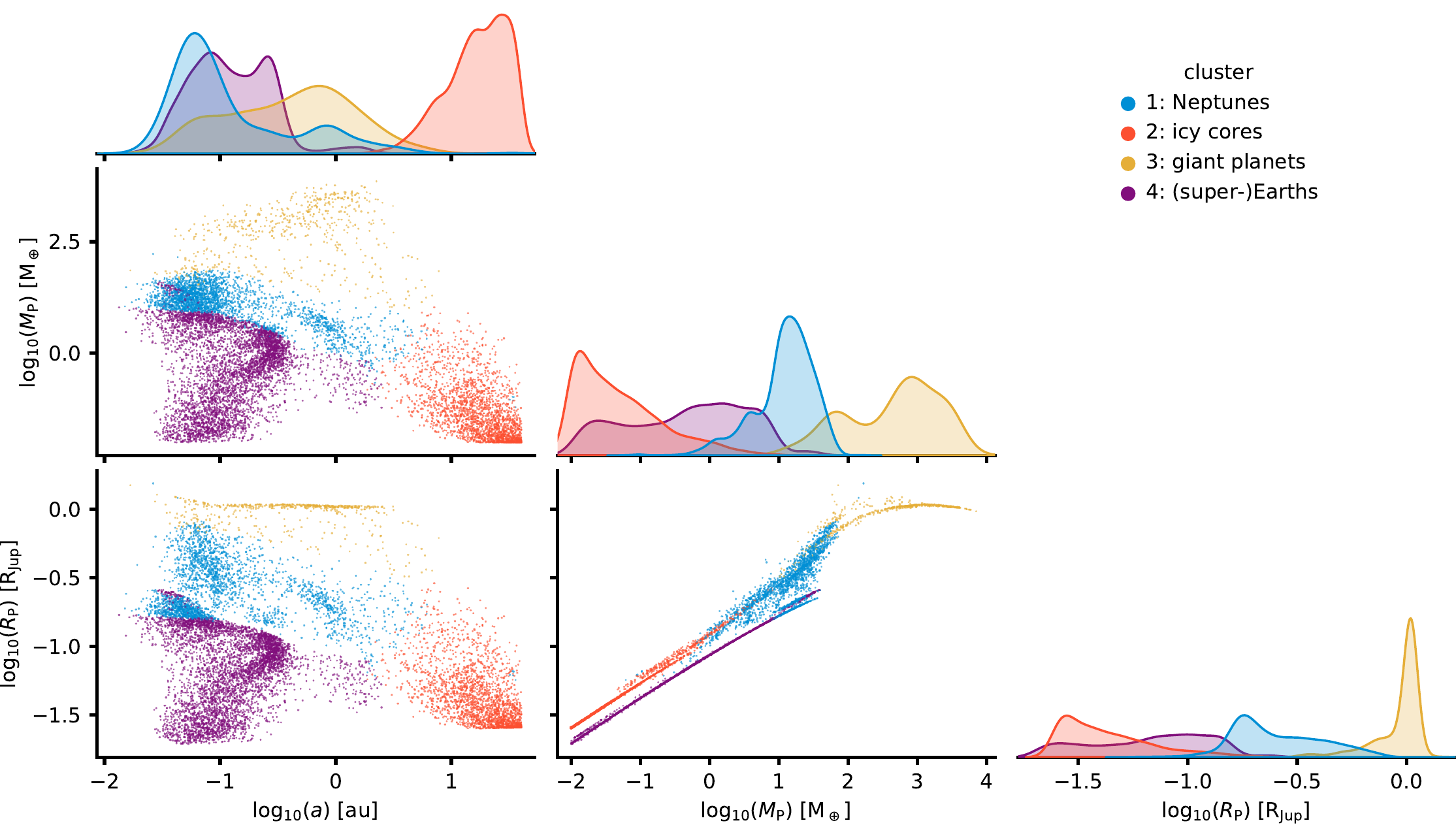}
	\caption[Single-planet population clustering.]{Planet clusters in a \SI{5}{\giga\year} old synthetic planet population \rev{with a single planet per system}. For all combinations of planet observables $a$, $M_\mathrm{P}$, and $R_\mathrm{P}$, the different colors denote clusters identified by a four-component Gaussian Mixture Model (GMM).
	On the diagonal, we show Kernel Density Estimates of the distributions.
	Without any information about the physics in our formation model, the GMM identified four planet species roughly corresponding to \mbox{(sub-)Neptunes} (blue), icy cores (red), giant planets (yellow), and (super-)Earths (purple).
	}
	\label{fig:gmm_NG73}
\end{figure*}
\begin{figure*}
	\centering
	\includegraphics[width=\hsize]{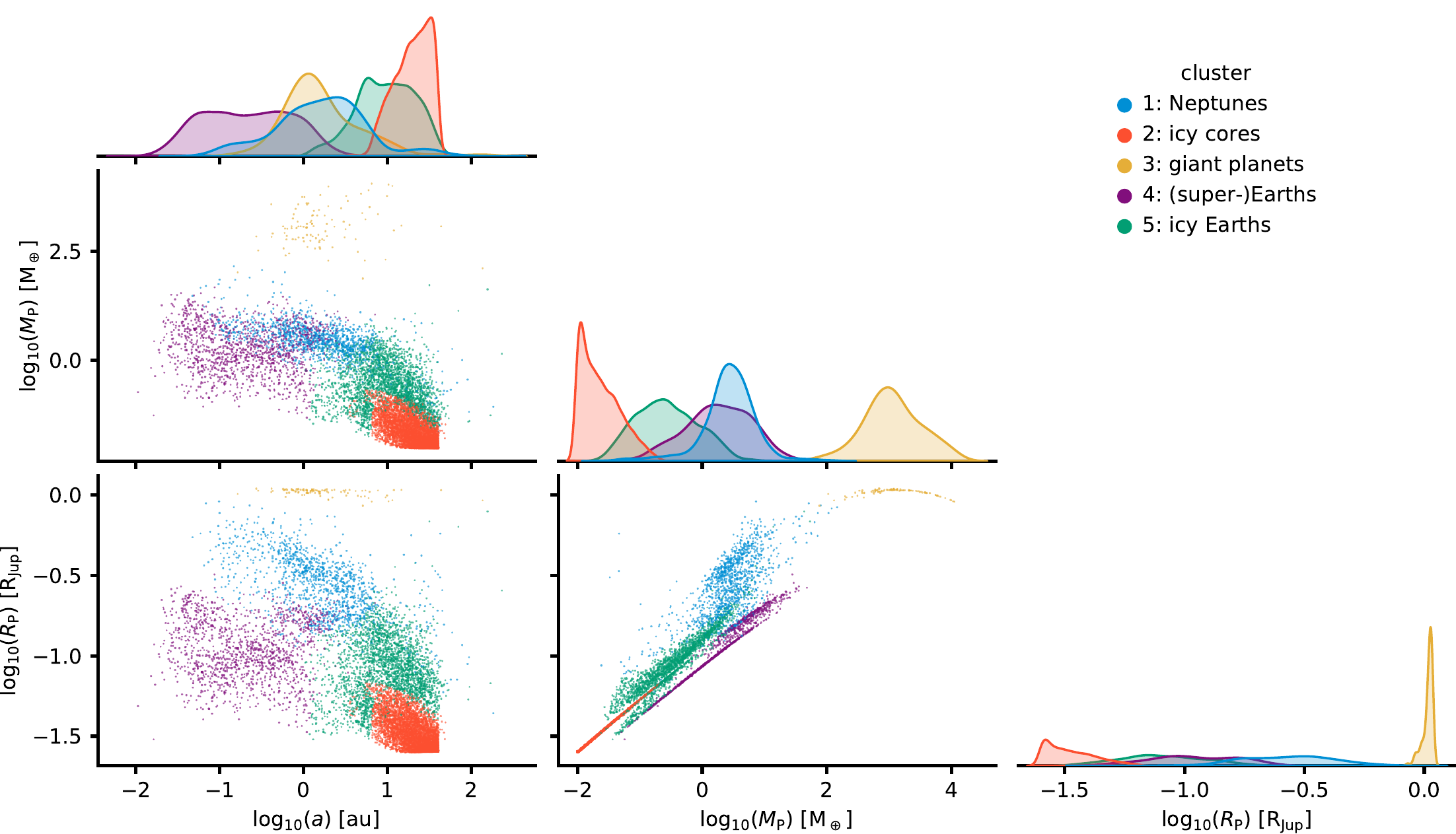}
	\caption[Multi-planet population clustering.]{Same as Fig.~\ref{fig:gmm_NG73}, but for a multi-planet population. The Gaussian Mixture Model (GMM) prefers solutions including a fifth component of distant, icy planets shown in green. In general, the clusters are less clearly separated than in the single-planet population.
	}
	\label{fig:gmm_NG76}
\end{figure*}
\rev{In the single-planet case, } the clustering algorithm identified four separate planet species in our population.
Figure~\ref{fig:gmm_NG73} shows these clusters in the various projections in $\{a, M_\mathrm{P}, R_\mathrm{P}\}$ space.
In general, we notice clear separations between the clusters in all projections, albeit with visible contaminations.
Ordered by ascending planetary mass, the clusters are as follows:
clusters~2 and 4 are populations of bare planet cores without atmospheres, and they are cleanly separated in semi-major axis.
Both clusters are separate from cluster~1, which are close-in planets enhanced in gas and with masses of mostly tens of \SI{}{\mEarth}.
A forth distinct group of very massive planets ($M_\mathrm{P} \gtrsim \SI{100}{\mEarth}$) is formed by cluster~3 with a clear separation from the other species.

Since the GMM is not aware of the underlying physics these clusters result from, it is of interest to interpret the identified clusters and relate them to known planet types.
Cluster~2 corresponds to an unobserved population of distant, low-mass planets.
As they formed beyond the water ice line and are rich in volatile species, we refer to this group as ``icy cores''.
Cluster~4 planets are atmosphere-less and rocky, and thus comparable to the observed population of close-in terrestrial planets and super-Earths\rev{~\citep[e.g.,][]{Hsu2019}}.
By simultaneously taking into account all dimensions of the parameter space, the GMM spatially separated icy cores and (super-)Earths in the region of the water ice line (without any information about its existence).
	This lead to the clean separation of rocky and icy planets in the $M_\mathrm{P} - R_\mathrm{P}$ diagram (diagonal lines in the plot).
Cluster~1 roughly corresponds to the observed population of (sub-)Neptunes.
	In planet radius space, these planets are mostly located above the radius valley~\citep[e.g.,][see discussion in Sect.~\ref{sec:predict_NG76_results}]{Fulton2017,Mordasini2020}.
	There is some contamination by cluster~1 planets in the region of the largest and closest super-Earths, which we attribute to the inability of a GMM to fit a deviation from the otherwise extremely straight line of cluster~4 planets in $M_\mathrm{P} - R_\mathrm{P}$ space.
Finally, cluster~3 can be identified as gas giant planets.
This becomes especially clear in the $M_\mathrm{P} - R_\mathrm{P}$ plane, where they occupy the region where in the physical model electron degeneracy occurs.
This effect flattens off the mass-radius relation at the high-mass end~\citep[e.g.,][]{Chabrier2009}.

\rev{Figure~\ref{fig:gmm_NG76} illustrates the clustering in the multi-planet case, during which we ignored the system affiliation of the planets and treated them as independent entities.
Based on the scoring scheme described above, the clearest clustering can be achieved with five components.
The overall partitioning appears similar to before, and the fifth component not present in the single-planet population covers planets on distant orbits that have intermediate densities and masses of roughly \SIrange{0.05}{3}{\mEarth}.
	We refer to these planets as ``icy Earths''.
	These planets are distributed in a sharp line in mass-radius space, which makes the GMM consider them detached from the more dispersed ``icy cores''.
	Notably, the bulk of the ``Neptunes'' moved to more distant orbits compared to the single-planet case.
This is in line with the observed existence of Neptune-sized planets at orbital distances of several au~\citet{Suzuki2016,Kawahara2019}.
For a comparison of Bern model planets and gravitational microlensing events, we refer to~\citep{Suzuki2018}.
}

\subsection{\rev{Model validation}}
	\begin{figure*}
		\centering
		\includegraphics[width=\hsize]{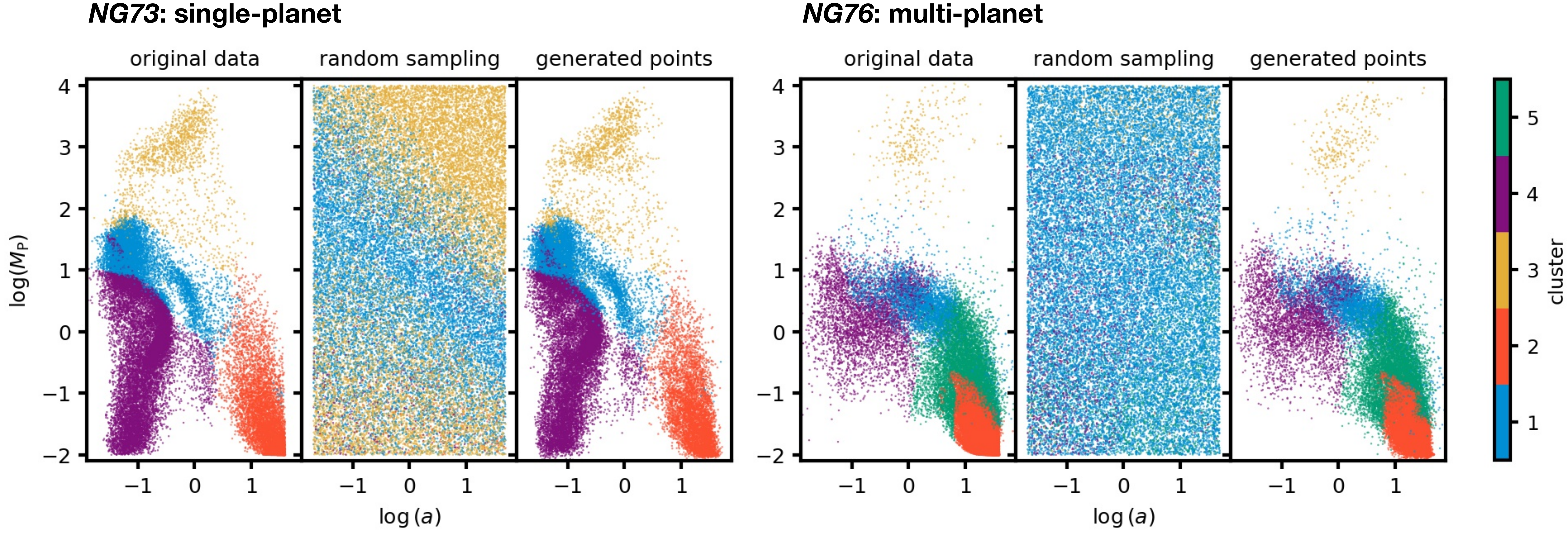}
		\caption{\rev{Model validation via generative models. For each of the two planet populations, we show the clustering result of our Gaussian Mixture Model on Population Synthesis data (left), random noise (center), and data from a generative model (right).
		Note that the latter do not stem from a physical formation model but were generated from a high-order GMM that was trained on the original data. }
		The clusters detected in these new data show largely the same structure as the original ones, whereas in the random noise no reliable clusters are found.}
		\label{fig:GMMvalidGenerated}
	\end{figure*}
	Unlike supervised machine learning algorithms, unsupervised techniques cannot be tested by applying the trained model to a test set due to the lack of "labeled" data. For validation of the clustering itself, we used the aforementioned performance metrics. To evaluate how robust the detected clustering is, we let the model predict the cluster affiliation of a data set of similar structure and compared these predictions to the original clustering. To produce these test data, we employed \rev{Gaussian Mixtures of 80 components and full covariance matrices as generative probabilistic models.
We trained them on the  $\{a, M_\mathrm{P}, R_\mathrm{P}\}$ subspace of the original population synthesis data.
The samples drawn from these models show a very similar structure in the whole domain (compare Fig. \ref{fig:GMMvalidGenerated}).}
Note that these ``planets'' are entirely the product of \rev{the generative models} and have never been in contact with a physical formation model.

	For comparison, we also fed \rev{our nominal clustering models with samples drawn from log-uniform distributions with boundaries roughly corresponding to the suprema of the population synthesis data, i.e. $a \sim 10^{\mathcal{U}(-1, 2)},\, M_\mathrm{P} \sim 10^{\mathcal{U}(-2, 4)},\, $and $R_\mathrm{P} \sim 10^{\mathcal{U}(-2, 0)}.$ With these pseudo-random data, the models predict clusters that do not resemble the original structures} and they appear in most projections almost random.
	These two tests show that our trained \rev{models neither overfit the data set, nor do they} produce any clear clusters where none are expected.
\rev{The generative models can also be used to draw a virtually unlimited number of synthetic planets when the computational costs of employing the full formation model are prohibitive~\citep[similar to][]{Mulders2018a}.}

\subsection{\rev{Planet clustering as a function of simulation time}}
    The cluster analysis took place at a simulation time of $t = \SI{5}{\giga\year}$.
	We now trace the identified clusters back in time to investigate their past evolution.
	\begin{figure*}
		\centering
		\includegraphics[width=\hsize]{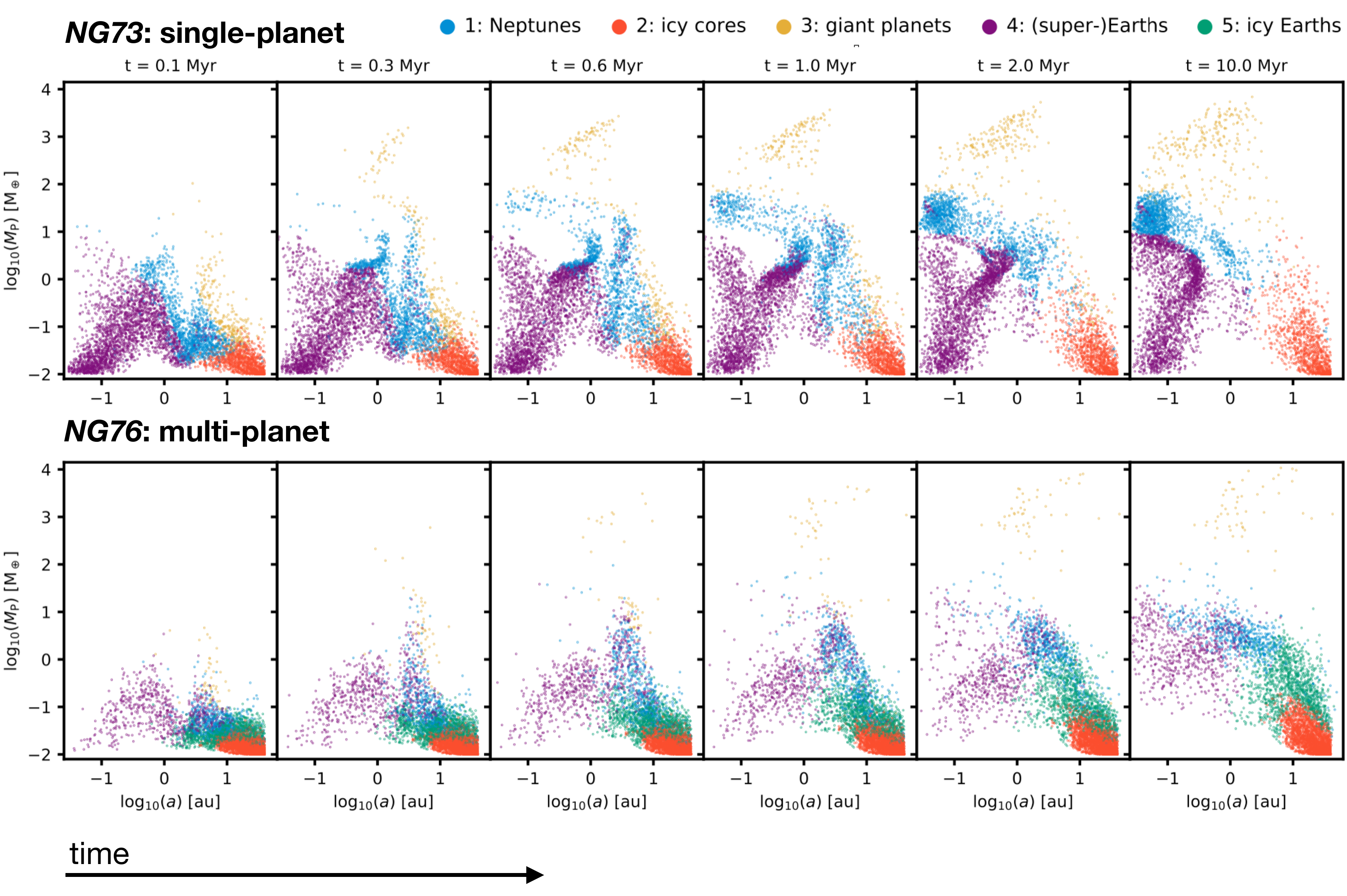}
		\caption{Early time evolution of the clusters identified by the Gaussian Mixture Model.
		Each subplot shows a sample of 5000~planets at their current position in semi-major axis-mass space and color-coded by their future cluster affiliation, which is only determined at $t = \SI{5}{\giga\year}$.
		Concurrent accretion and migration leads to characteristic evolutionary paths.
		Distinct groups of planets form already at early simulation times.
		}
		\label{fig:traceback}
	\end{figure*}
	Figure~\ref{fig:traceback} shows their position in semi-major axis-mass space at simulation times $\SI{0.1}{\mega\year}$, $\SI{0.3}{\mega\year}$, $\SI{0.6}{\mega\year}$, $\SI{1}{\mega\year}$, $\SI{2}{\mega\year}$, and $\SI{10}{\mega\year}$.
\rev{In particular in the single-planet population, the} clusters occupy distinct domains already at early times and follow characteristic paths in this parameter space.
	These paths are set by concurrent accretion and planet migration and their respective timescales.

	\rev{In the following, we focus on the single-planet case where the evolutionary paths can be traced most clearly.}	
	At the beginning, all planets are still of such low mass that migration has little effect.
	Planet growth is determined by the local planetesimal density, feeding zone size, and orbit timescale, and it is most efficient at intermediate orbital distances~(\papertwo).
	At a few $\SI{e5}{\year}$, an outward migration zone located at a few au divides the planetary tracks into two branches.
	On the outer branch, giant planets evolve similarly as the outer wing of Neptunes.
	They branch off when runaway gas accretion sets in, while Neptunes continue migrating inward with moderate growth.
	At later times, another outward migration zone leads to the underdensity in the cluster of close-in (super-)Earths.
	Icy cores do not exhibit significant growth and largely remain in their initial domain.

	Most of the processes that define the different planet types in this parameter space are finished after a few \SI{}{\mega\year} or, at the latest, when the gas disk disperses.
	Exceptions are atmospheric photoevaporation, which happens on $\SI{100}{\mega\year}$ to \SI{}{\giga\year} timescales~\citep[e.g.,][]{Lopez2012,King2020} and still turns some close-in (sub-)Neptunes into super-Earths, and tidal interaction with the host star affecting some ultra-short period planets.
	In the case of multiple planets per system, N-body interactions can have an additional long-term impact.
\rev{A striking result of planet-planet interactions are the significantly lower migration rates compared to the single-planet case, in particular in the Neptunes cluster.}

\rev{In general, it appears that planet populations form} distinct groups very early in the formation process.
	This begs the question whether the cluster affiliation of a planet can already be predicted from the initial conditions of the simulation.

\section{Prediction of planet clusters}\label{sec:prediction}
Our planet formation model provides a deterministic link between properties of protoplanetary disks and properties of planets.
\rev{This link could be blurred by N-body interactions between the planets, hence in the following experiment we consider first the single-planet population.
Our approach was to employ} a Random Forest classifier \citep{Ho1998, Breiman2001} to predict the cluster of a planet from its corresponding set of disk properties. %
	Random Forests are ensembles of uncorrelated, binary classifiers known as decision trees. 
	Such ensembles achieve strongly improved generalization accuracies compared to single-tree classifiers by constructing trees in pseudorandomly selected feature subspaces \citep{Ho1995}. 
	The individual trees are further decorrelated by drawing, with replacement, random subsets of the input data during training (``bagging'',  \citealp{Breiman1996}).
	
	With varying sizes of the individual clusters (for instance, only $\sim \SI{5}{\percent}$ of \rev{the planets in \textit{NG73}} are giant planets), our data set is strongly imbalanced.
	This is problematic for classification algorithms such as Random Forests, which aim to minimize the overall error rate and thereby tend to neglect minority classes~\citep{Chen2004}.
    To account for this imbalance, we employed a balanced Random Forest classifier as implemented in the imbalanced-learn\footnote{https://imbalanced-learn.org} python package.
This variant of Random Forest randomly under-samples each bootstrap sample on the individual tree level during training~\citep{Lemaitre2017}.

\subsection{Data preparation, hyperparameters, and training}\label{sec:dataprep}
	Our classifier learned rules based on four features: the initial gas disk mass $M_\mathrm{gas,0}$, the initial solid disk mass $M_\mathrm{solid,0}$, the initial orbital distance of the planetary embryo $a_\mathrm{start}$, and the disk lifetime $t_\mathrm{disk}$.
The solid disk mass is a derived quantity that we computed from the gas disk mass and host star metallicity.
	We rescaled these features to account for their large differences in scale: $t_\mathrm{disk}$ and $a_\mathrm{start}$ were transformed by a $\log_{10}$ function, and $M_\mathrm{gas,0}$ and $M_\mathrm{solid,0}$ were modified to roughly Gaussian distributions by the Box-Cox transform \citep{Box1964}.
The clustering above assigned each synthetic planet a probability to belong to each of \rev{the clusters}.
	For the subsequent analysis, we avoided planets that cannot be mapped clearly to a cluster and kept only those with a probability of affiliation $>0.99$.
	This decreased our sample from 29455 to 23278 planets.
	Finally, we divided the data into a random subset containing 80\% of the initial data for training and a test set with the remaining 20\% to determine the performance of the classifier.
	The resulting training set contains between 1059 (giant planets) and 8486 ((super-)Earths) planets per cluster.
	We trained an ensemble of 500 fully grown estimators, that is, without reducing the depth of the trees by pruning them, on this set.

\subsection{Error and performance analysis}\label{sec:error_performance}
	To measure the generalization performance of the trained model already during its development, we predicted clusters from the out-of-bag samples, which were never seen by the respective estimator during training. The average of the resulting out-of-bag score produces an estimate for the accuracy of the entire ensemble, and we obtained a score of 98\% here.
	However, classification accuracy is not a sufficient performance measure since we are dealing with a strongly skewed data~set.
In the following, we investigate the types of errors our model makes and measure its performance.

\begin{figure*}
	\centering
	\includegraphics[width=\hsize]{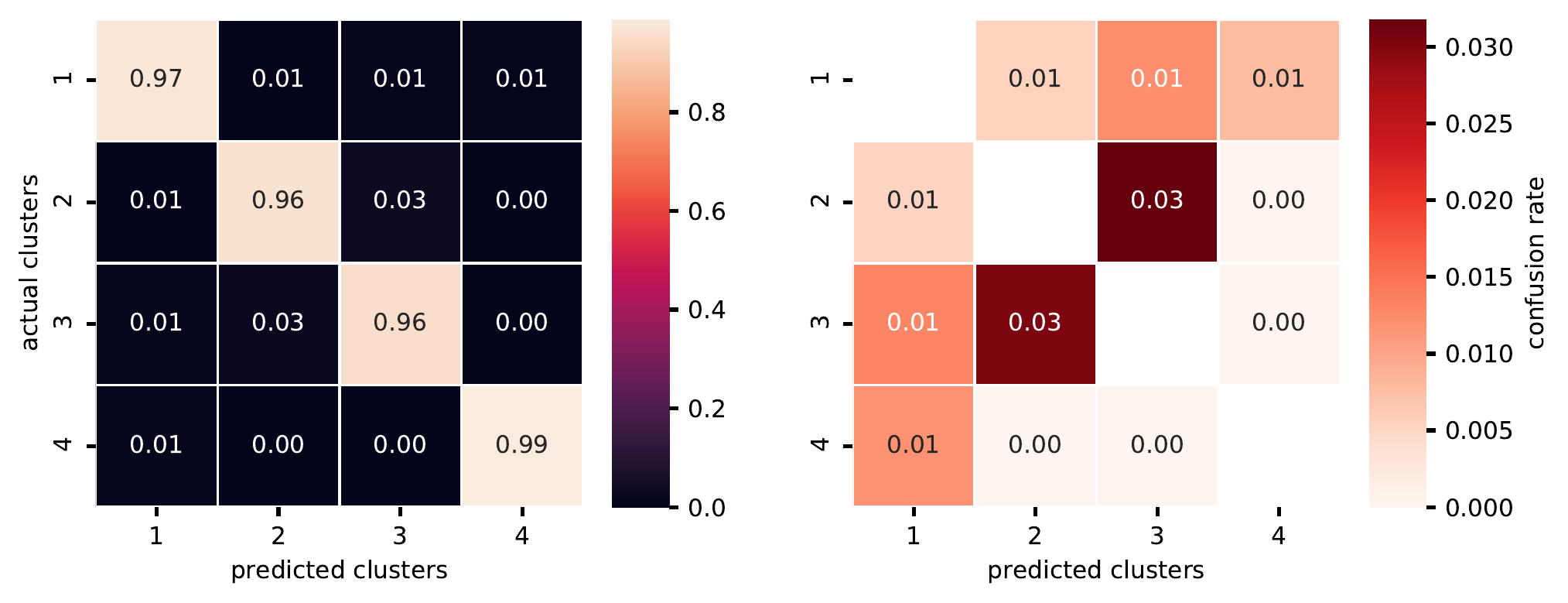}
	\caption[Confusion matrix of the planet classification.]{Confusion of planet classifications. Left: confusion matrix from five-fold cross-validation. Rows are the actual clusters and columns are the predicted clusters.
	All clusters are classified with more than 95\% accuracy. Right: same, but correct classifications removed to emphasize errors. Most misclassifications occur between clusters 2 and 3,
	which correspond to icy cores and giant planets.
	}
	\label{fig:rfconfusionmatrix}
\end{figure*}

	We computed a confusion matrix using five-fold cross-validation. For this purpose, the data set was randomly split into five evenly sized folds; the model was trained five times on $5-1=4$ folds, and then evaluated on the fold it was not trained on. 
	The left panel of Fig. \ref{fig:rfconfusionmatrix} shows the confusion matrix produced from the labeled training set and the predictions from cross-validation. Rows correspond to the actual clusters, and columns are the predictions of our model. Each field $x_{i,j}$ in the matrix shows the fraction of times a planet of cluster $i$ was classified as a planet of cluster $j$. Most planets fall into the diagonal, meaning a correct classification.
	All clusters are predicted with more than 95\% accuracy and the largest errors occur for clusters 2 and 3.
	The right panel of the figure shows the same matrix with the correct classifications removed and the color map rescaled.
	It is obvious that the errors are largely symmetric.
	The highest rate of misclassification occurred between clusters 2 and 3 (3\% of icy cores were confused with giant planets and vice-versa).
    The reason is that the former are frequently progenitors of the latter, and prediction of those planets that just (do not) reach the conditions for runaway gas accretion is difficult (compare Fig.~\ref{fig:gmm_NG73}).

	To estimate the generalization error the model makes when applied to data not part of the training set, we measured its performance on the test set of 4656 systems we held out before.
	Based on five-fold cross-validation, it achieves an overall accuracy of \SI{97}{\percent} and misclassifications occur between the same clusters as seen in the training set.
	This shows that the model is not significantly overfitted.

\subsection{Results of planet predictions}
	\subsubsection{Correlations with disk properties}\label{sec:diskprops}
\begin{figure*}
		\centering
		\includegraphics[width=\hsize]{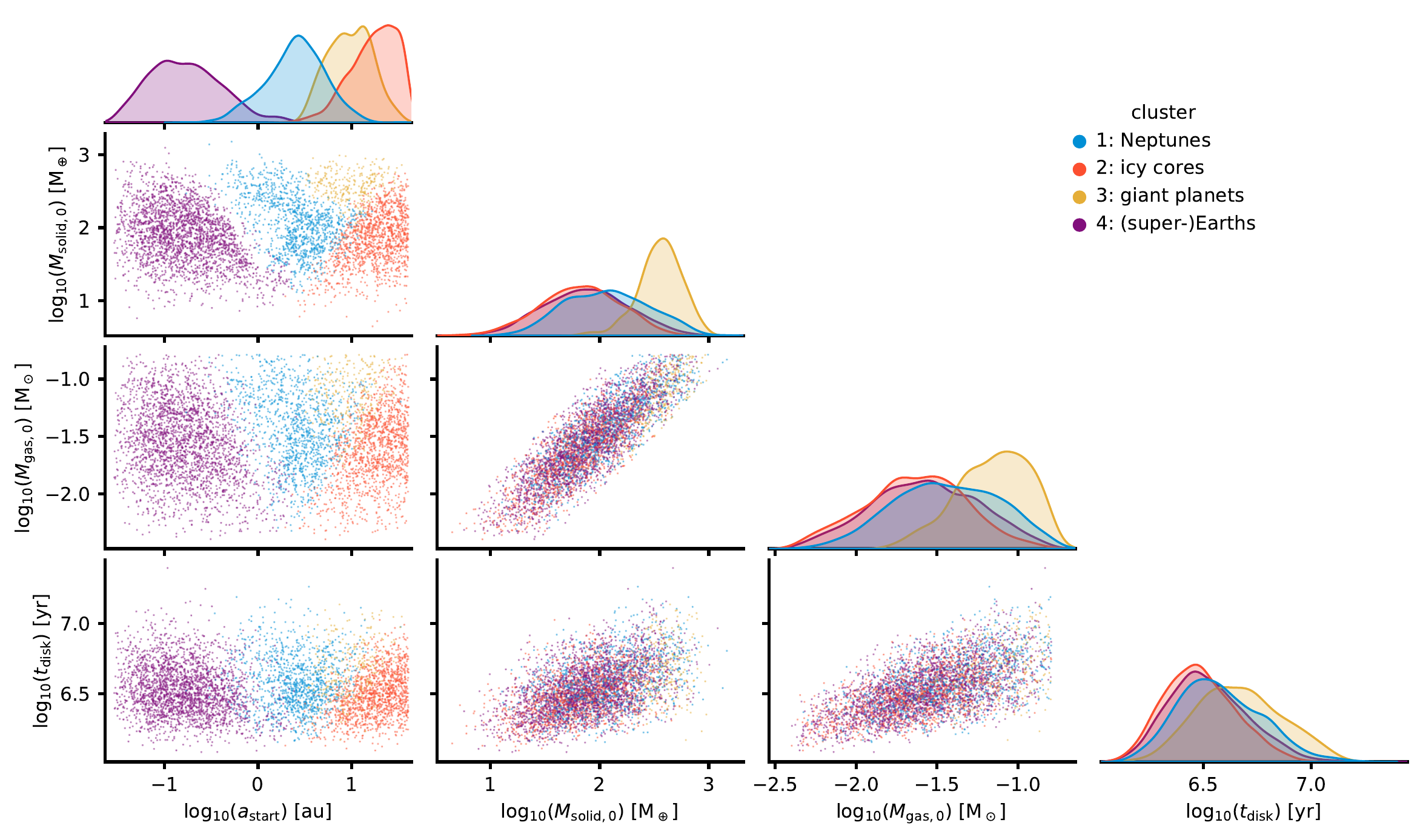}
		\caption{Pairwise relationships between all disk parameters, sorted by cluster affiliation. For 5000 randomly sampled planets in the population, each parameter is plotted against every other parameter while the color defines the planet's cluster. The diagonal panels show the univariate distributions of the respective parameters, again colored by cluster assignment.
		Planet species most clearly separate in $a_{\mathrm{start}} - M_\mathrm{solid,0}$ space, and the formation of giant planets (yellow) requires large solid reservoirs and a narrow range of initial orbital distance.}
		\label{fig:InitialsPairplot}
	\end{figure*}

	For each of the clusters identified in Sect.~\ref{sec:results:planetclusters}, we show the distributions and pairwise relationships of their corresponding disk properties in Fig. \ref{fig:InitialsPairplot}.
	Underdensities in the scatter plots are due to removed planets of ambiguous cluster affiliation.
	Unsurprisingly, giant planets (yellow) grow in disks with large reservoirs of solid material $M_{\mathrm{solid}}$ and high gas mass $M_{\mathrm{gas}}$.
	It is evident that most of these clusters, which are labeled at ``observation time'' $t = \SI{5}{\giga\year}$, form groups already in this parameter space, that is, before the simulations started.
	However, they differentiate distinctly only in the projections involving the start position of planetary embryos $a_{\mathrm{start}}$.
	The separation is especially clear in $a_{\mathrm{start}} - M_\mathrm{solid,0}$ space, which shows the least overlap of different clusters.
	With increasing initial orbital distance, the dominant planet species are (super-)Earths, Neptunes, giant planets, and icy cores.

	\subsubsection{Feature importance}\label{sec:featureImportance}
	
	Our classification model reaches high accuracies for all planet clusters, but it is interesting to see which disk features are most important for a successful classification.
	This is possible by measuring the feature importance of the data set given to the model using the Mean Decrease Impurity $MDI$~\citep{Breiman1984}.
	$MDI$ quantifies to what extent a feature reduces the impurity of the trees in the Random Forest.
	Put simply, it is a measure of how well the nodes can use the feature to split the data~set into ``pure'' child nodes, each containing only data of a single label.
	A higher score means that the feature is more important for correct classification.
	We list the $MDI$ for each input parameter in Table~\ref{tab:featureImportance}.
	\begin{table}
		\caption{Feature importances of disk properties}             %
		\centering                          %
		\label{tab:featureImportance}
		
		\begin{tabular}{r c c c c}        %
			\hline\hline                 %
			Input Parameter & $M_\mathrm{solid,0}$ & $M_\mathrm{gas,0}$ & $a_\mathrm{start}$ & $t_\mathrm{disk}$\\    %
			\hline                        %
			$MDI$ & 0.21 & 0.07 & 0.68 & 0.04 \\      %
			
			\hline                                   %
		\end{tabular}
	\end{table}
	With a score of $0.68$, the starting position of the planetary core $a_{\mathrm{start}}$ is clearly the parameter most sensitive for predicting a planet's cluster.
	The gaseous mass of the disk and its lifetime are the least important features.
	
\begin{figure*}
		\centering
		\includegraphics[width=\hsize]{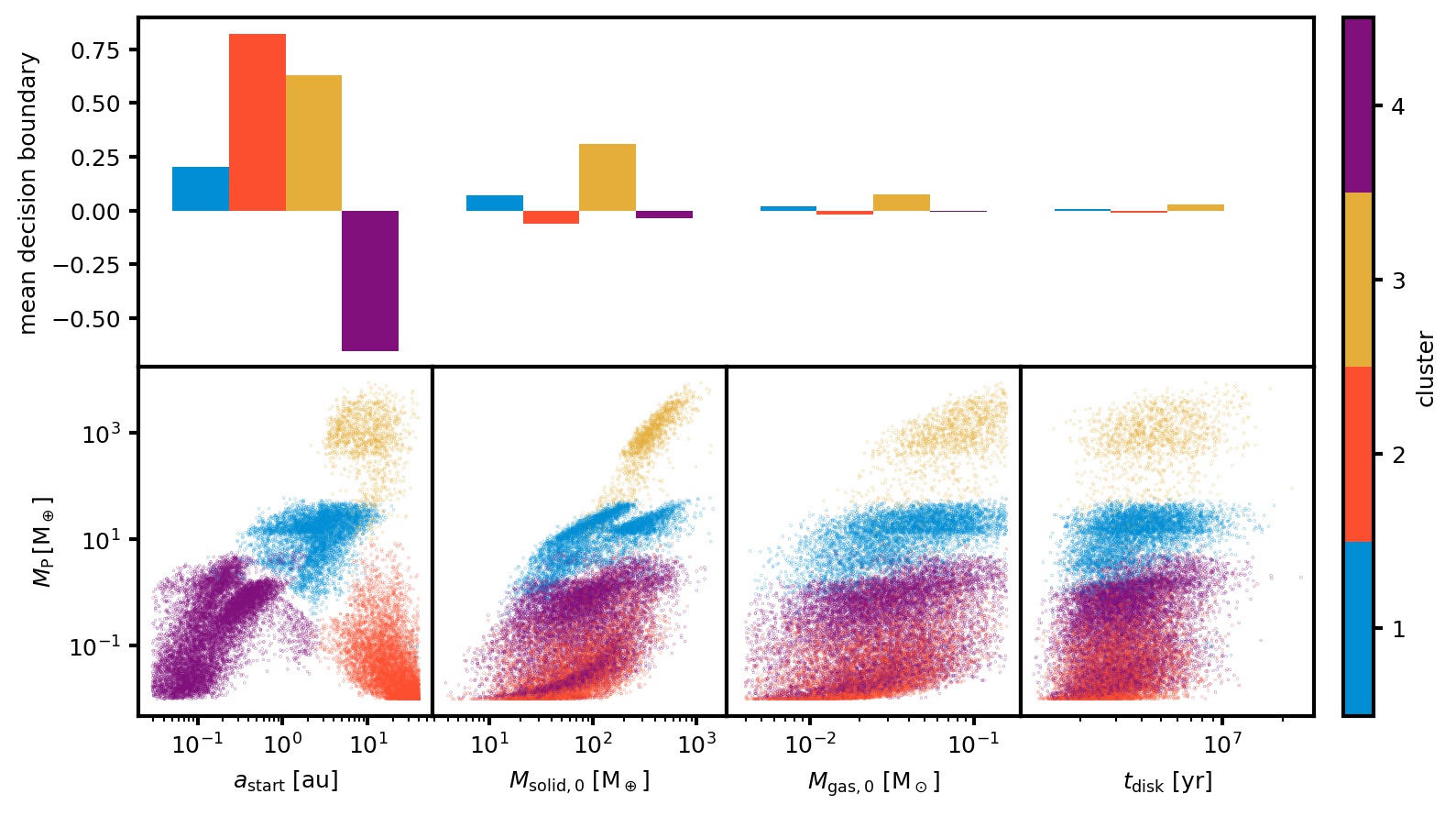}
		\caption{Relation between disk features and planet species.
		Upper panel: Mean decision boundaries of the classifier, indicating the importance of each feature and its preferred magnitude for the different clusters.
		The starting location of the planet embryo $a_\mathrm{start}$ shows the largest variance in decision boundary.
		Giant planets (yellow) are also very sensitive on $M_\mathrm{solid,0}$ and somewhat sensitive on $M_\mathrm{gas,0}$.
		Lower panels: relationship of the input features with planet mass.
		The starting location of the planet embryo $a_\mathrm{start}$ shows the strongest correlation with cluster affiliation and planet mass.
		}
		\label{fig:initial-mass_per_cluster}
	\end{figure*}

	However, the degree of dependency on certain disk features varies from cluster to cluster.
	To get a cluster-specific insight, we multiply for each cluster the mean of each feature with the feature importance.
	This mean decision boundary
	\begin{equation}\label{eq:meanDecisionBoundary}
	D_{c,f} = MDI_f \cdot \left<X_{y=c}\right>
	\end{equation}
	denotes for each cluster $c$ the sensitivity of the classifier on feature $f$.
	Here, $X_{y=c}$ are the scaled training data with labels $y$ corresponding to cluster $c$.
	Fig.~\ref{fig:initial-mass_per_cluster} illustrates all cluster-specific mean decision boundaries.
	$D_{c,f}$ quantifies the sensitivity on a parameter by its magnitude, as well as the orientation of the decision boundary by its sign.
	For example, the large negative value of cluster 4 in $a_\mathrm{start}$ means that these planets prefer small initial orbital distances and their correct classification is very sensitive on this feature.

	In the lower panels of Fig.~\ref{fig:initial-mass_per_cluster}, we plot all input features against the resulting planet mass at \SI{5}{\giga\year}, which is a proxy for cluster affiliation.
	Most planet clusters are especially sensitive on the initial orbital distance of the planetary embryo $a_\mathrm{start}$.
	Planets with masses higher than $\sim \SI{10}{\mEarth}$ are also very sensitive on the solid mass $M_\mathrm{solid,0}$ and slightly sensitive on $M_\mathrm{gas,0}$.
	The disk lifetime $t_\mathrm{disk}$ shows a weak correlation with planet mass and plays only a subordinate role.
	
\subsection{\rev{Differences between single and multi-planet systems}}\label{sec:predict_NG76_results}
\rev{Mutual interactions between planets in the same system introduce} a fair amount of stochasticity, and some features that stood out in the single-planet population are \rev{smeared out in the multi-planet case}.
	One example is the bimodal distribution of planet radii in the observed exoplanet sample~\citep{Fulton2017,Fulton2018,Hsu2018,VanEylen2018,Mordasini2020}, which was theoretically predicted to be caused by photoevaporation of planetary envelopes by high-energy radiation from their host star~\citep{Jin2014,Owen2013,Lopez2013}.
Other mechanisms have been proposed to produce this ``radius valley'' at roughly \SI{2}{\rEarth} as well, including atmospheric loss due to internal heat from cooling planetary cores~\citep{Ginzburg2017,Ginzburg2018,Gupta2019}, impacts of planetesimals~\citep{Wyatt2019a} or other protoplanets~\citep{Liu2015}, different internal compositions of planets residing above or below the valley~\citep{Zeng2019,Venturini2020a}, and atmospheric stripping by external radiation sources in stellar cluster environments~\citep{Kruijssen2020}.
In the Generation III \textit{Bern} Model, photoevaporation by the host star and collisional stripping are taken into account.

\begin{figure*}
	\centering
	\includegraphics[width=\hsize]{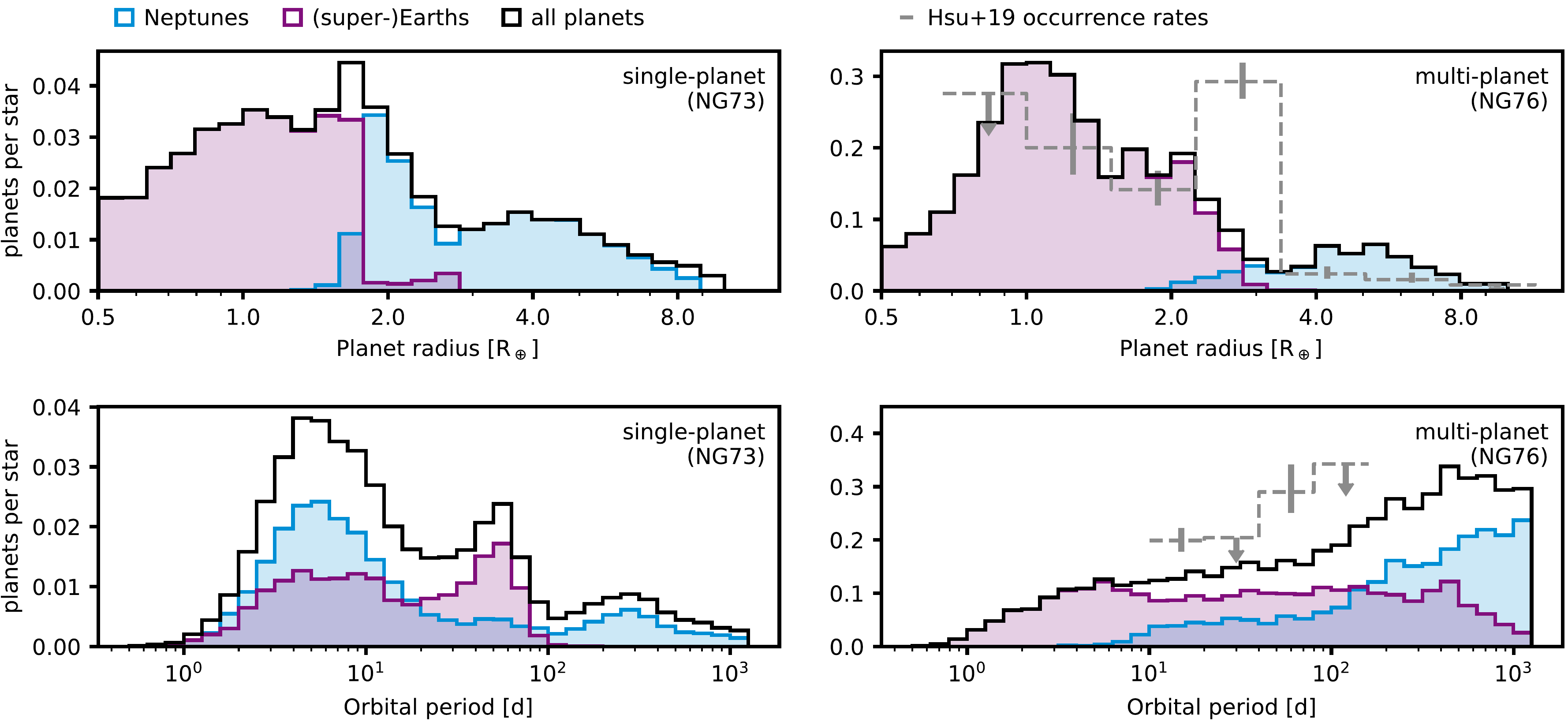}
	\caption[Radius and period distributions of Neptunes and (super-)Earths]{\rev{Radius and period distributions of Neptunes and (super-)Earths. The contributions by Neptunes and (super-)Earths are shown in blue and purple, respectively. Upper panels: planet} radius distribution for planets with periods \rev{$P < \SI{80}{\day}$}. \rev{In the single-planet case (left)}, a population of migrated, icy cores in the Neptunes cluster shifts the synthetic radius valley to larger radii. \rev{In the case of multiple planets per system (right)}, the minimum in the distribution separates (super-)Earths and Neptunes. \rev{Compared to observed occurrence rates from \textit{Kepler}~\citep[][gray]{Hsu2019}, this minimum is shifted towards larger radii.\\
Lower panels: period distributions of planets $\geq \SI{1}{\rEarth}$. While the single-planet population (left) shows a multi-modal distribution, the multi-planet population has a continuous slope similar to observed occurrence rates.
	Note the different normalizations of synthetic and observed planets.}}
	\label{fig:radiusValley}
\end{figure*}
\rev{The upper panels of Fig.~\ref{fig:radiusValley} show the radius distributions} of planets on close orbits (\rev{$P < \SI{80}{\day}$}) in the single and multi-planet populations, respectively.
\rev{
Overplotted are occurrence rates derived from the \textit{Kepler} mission in~\citet{Hsu2019}, which we marginalized over the period range \SIrange{0}{80}{\day}.
The propagated uncertainties are indicated by vertical bars, and arrows mark upper limits.
In our} single-planet population, the evaporation valley is much less pronounced in this marginalized radius distribution than in radius-orbital distance space, where it shows a steep negative slope (compare Fig.~\ref{fig:gmm_NG73}).
This highlights the importance of characterizing such demographic features in multiple dimensions.
Compared to the observed valley at $\sim \SI{2}{\rEarth}$\rev{~\citep[e.g.,][]{Fulton2017,Hsu2019}}, the synthetic one is shifted to larger radii.
As has been shown in~\citet{Jin2018}, this is due to atmosphere-less, icy cores that migrated inwards from regions beyond the water ice line.
This population is included in the planet cluster representing Neptunes, since the clustering algorithm mainly discriminated between (super-)Earths and Neptunes as rocky and icy planets, respectively.

In the multi-planet population, this is not the case.
Here, the different clusters divide close-in planets into bare cores and planets with H/He envelopes, and the emerging radius valley separates the (super-)Earths and Neptunes clusters.
Again, the valley is shifted to around \SI{3}{\rEarth}.
Compared to the single-planet case, the slope of the valley in radius-orbital distance is less pronounced, which makes it appear deeper in the one-dimensional radius histogram.
	Future work within this series will address the synthetic radius valley in a more thorough manner~(Mishra et al., in prep.).

\rev{Other differences between the single and multi-planet populations can be seen in their period distributions~(lower panels of Fig.~\ref{fig:radiusValley}).
In the single-planet case, the combined contributions from (super-)Earths and Neptunes lead to a multi-modal period distribution.
On the other hand, the multi-planet population shows a continuous slope.
In the range where \citet{Hsu2019} provide reliable occurrence estimates, this slope matches the observed one well.
Causes for the difference between the single- and multi-planet case are the displacement of planets in semi-major axis due to gravitational encounters, a lack of close-in ``failed cores'' due to the high likelihood of such encounters on short orbits, and trapping of planets in resonant chains.}
\rev{In addition, mixed planetary compositions occur as a consequence of merger events.
This places the planets into a continuum of bulk densities.}

\rev{Regardless of this ``stochastic processing'' of the planets, we attempted to predict their clusters from initial conditions} using the same features as in the single-planet case and following the procedure described in Sects.~\ref{sec:dataprep}~to~\ref{sec:error_performance}.
	Similar to before, keeping only planets that the GMM assigned to a specific cluster with a probability $> 0.99$ reduces the set to 21,761 planets.
The randomly drawn training set comprising $80\,~\%$ of the data contains between 252 (giant planets) and 10367 (icy cores) planets per cluster.
	A balanced Random Forest we trained on this set achieved an accuracy of 89\,~\% based on five-fold cross-validation.
	The other 4353 systems, which we left out as a test set, are predicted with \SI{86}{\percent} accuracy.

\begin{figure*}
	\centering
	\includegraphics[width=\hsize]{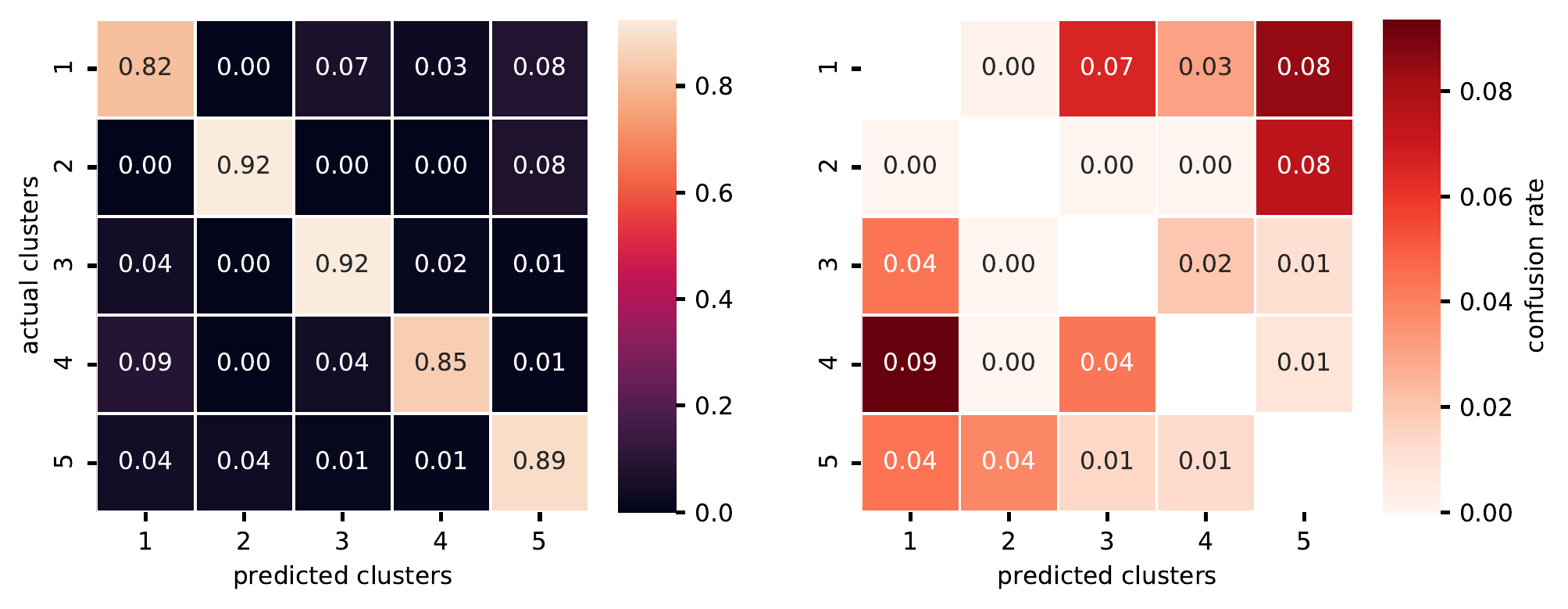}
	\caption[Confusion of classifications for N-body population.]{Confusion of cluster classifications for a multi-planet population with N-body interactions. Same as Fig. \ref{fig:rfconfusionmatrix}, but computed for a population with 100 planets per disk that interact gravitationally. Clusters 2 (icy cores) and 3 (giant planets) are predicted most reliably. Due to giant collisions the classifier cannot predict, the super-Earths in cluster 4 are often mistaken for (sub-)Neptunes (cluster 1).
	}
	\label{fig:rfconfusionmatrix_wNbody_balanced}
\end{figure*}
Similar to Fig.~\ref{fig:rfconfusionmatrix}, Fig.~\ref{fig:rfconfusionmatrix_wNbody_balanced} shows the confusion matrix of a Random Forest predicting the planet clusters in the multi-planet population. %
The ability to predict planet clusters from initial conditions varies across different planet types, with icy cores and giant planets being the most robust species.
It can be seen that clusters 1 (Neptunes) and 4 ((super-)Earths), which occupy similar mass ranges, are affected by confusion the most.
This is mainly due to the lack of (super-)Earths $\lesssim \SI{0.1}{\mEarth}$ in the multi-planet case, where they typically fall victim to giant collisions with other planets.
Neptunes are frequently mistaken as icy Earths and (super-)Earths are frequently confused to be Neptunes.
These three groups of intermediate-mass planets share a similar domain in parameter space.
\begin{figure*}
	\centering
	\includegraphics[width=\hsize]{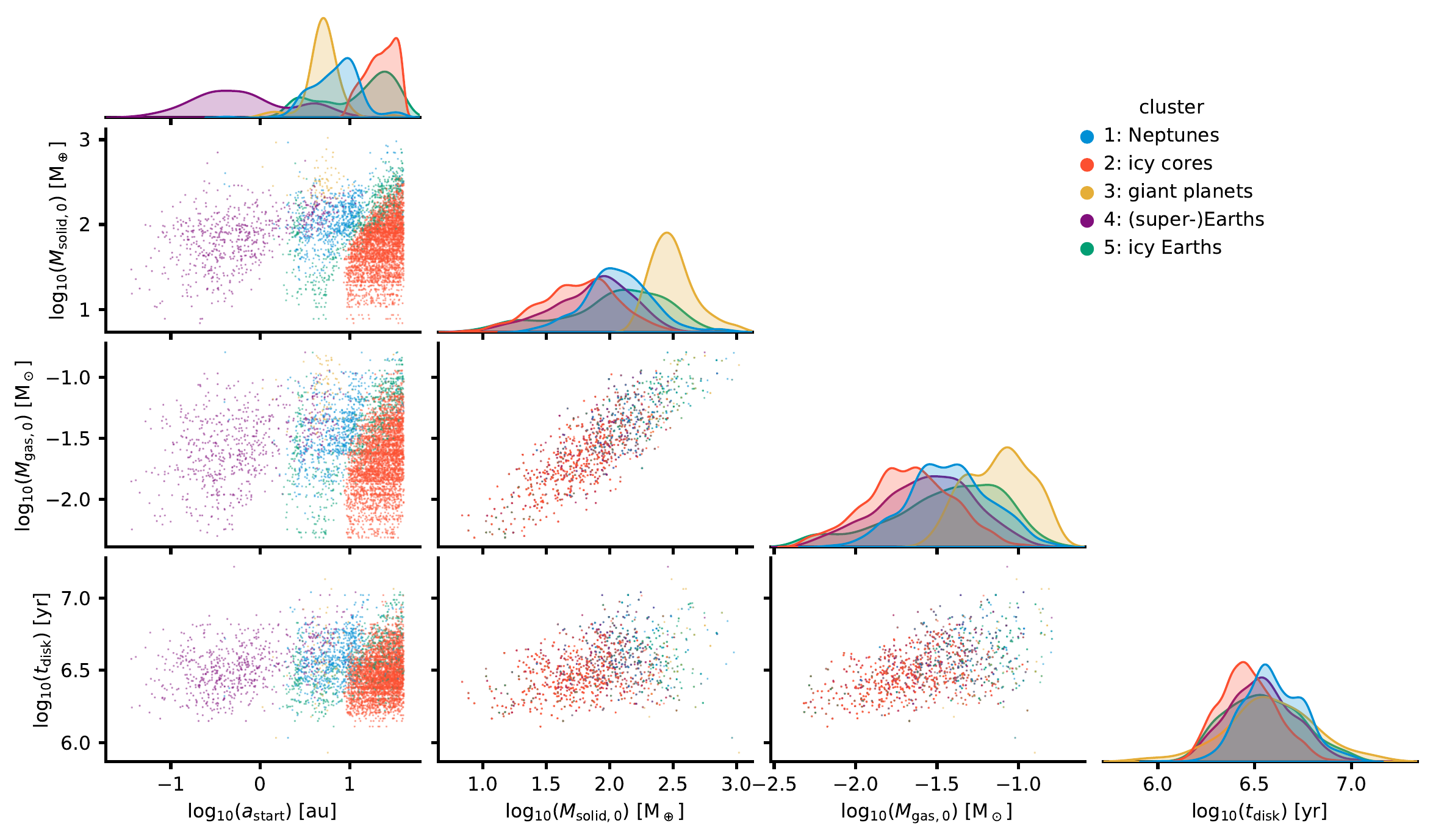}
	\caption[]{Pairwise relationships between all disk parameters, sorted by cluster affiliation. Same as Fig.~\ref{fig:InitialsPairplot}, but for a multi-planet population with N-body interactions. The separation of clusters is less pronounced than in the single-planet case.}
	\label{fig:InitialsPairplot_NG76}
\end{figure*}

Figure~\ref{fig:InitialsPairplot_NG76} shows the positions of the planets in the multi-planet population in disk property space.
Again, the different clusters differentiate the most in solid disk mass and initial orbital separation.
Compared to the single-planet case, the separation of the clusters is less clean.
The additional cluster identified in \textit{NG76}, ``icy Earths'', share a lot of parameter space with other planet types.
\begin{figure*}
	\centering
	\includegraphics[width=\hsize]{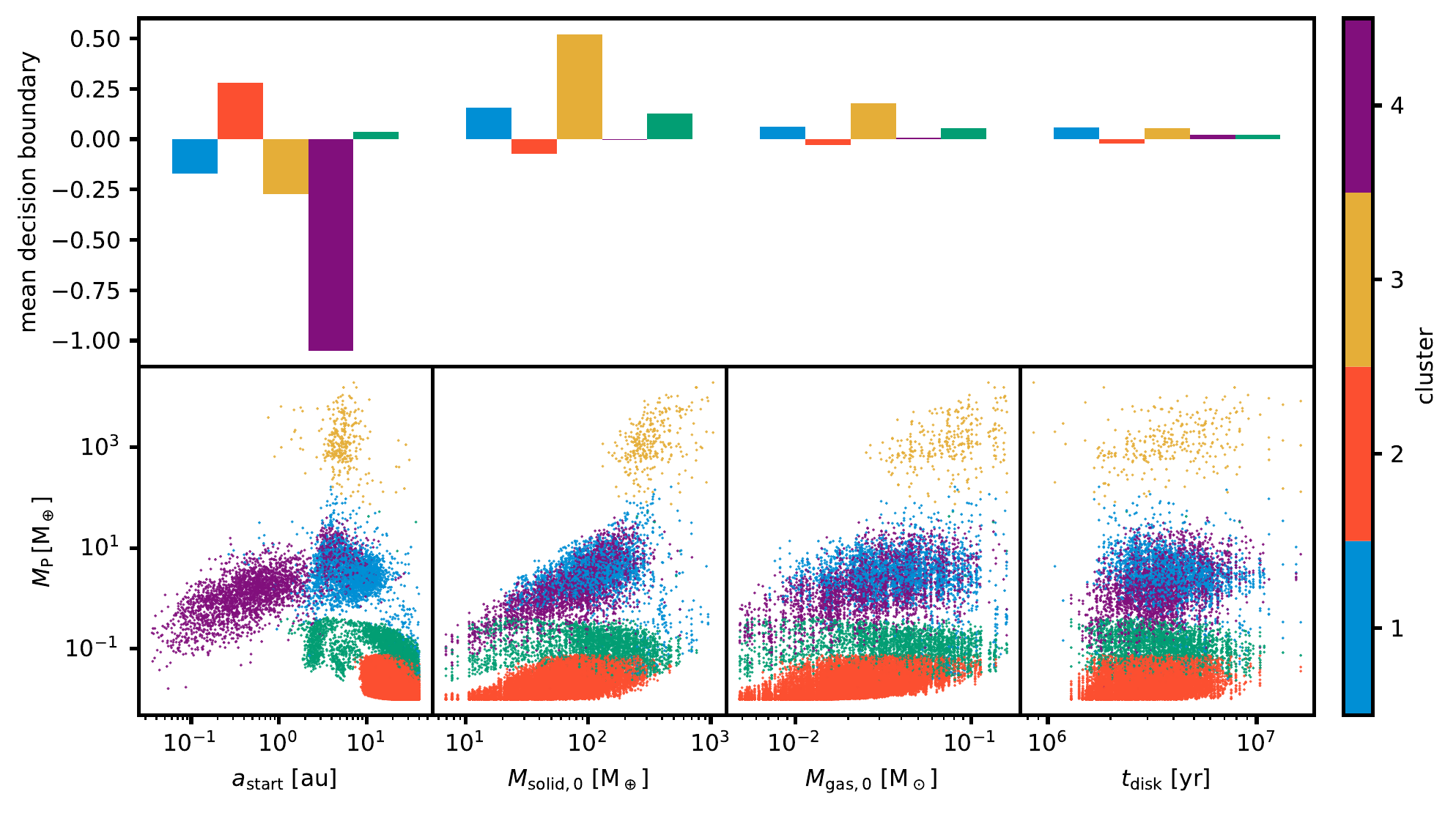}
	\caption{Relation between disk features and planet species. Same as Fig.~\ref{fig:initial-mass_per_cluster}, but for a multi-planet population with N-body interactions.
	As in the single-planet case, the starting location of the planet embryo $a_\mathrm{start}$ shows the largest variance in decision boundary.
	Giant planets (yellow) form only at high $M_\mathrm{solid,0}$ and sufficient $M_\mathrm{gas,0}$.
	}
	\label{fig:initial-mass_per_cluster_NG76}
\end{figure*}

Using the mean decision boundary defined above (Eqn~\ref{eq:meanDecisionBoundary}), the dependence of different planet clusters on specific initial conditions can be visualized also for the multi-planet population (Fig.~\ref{fig:initial-mass_per_cluster_NG76}).
The relationships largely copy those of the single-planet case: the starting location of the planet embryo shows the largest decision boundary amplitudes and differences among the clusters, and giant planets retain their distinct dependence on high solid and gas reservoirs.

\section{Discussion}\label{sec:discussion}

\subsection{What determines the type of a planet?}
By predicting a planet's cluster from a set of initial conditions of our planet formation model, we were able to establish links between properties of the protoplanetary disk and the corresponding planets (see Sect.~\ref{sec:featureImportance}).
	These links can be elucidated by using the planet mass $M_\mathrm{P}$ as a proxy for the planet cluster and relating it to disk features (see Fig. \ref{fig:initial-mass_per_cluster}).
	The feature with by far the highest predictive power is the starting location of the emerging protoplanetary embryo $a_\mathrm{start}$, which is expected in a core accretion scenario: 
	an embryo at small orbital distance has only a small feeding zone from which it can accrete and thus it will remain small.
	At very large orbital distance, the dynamical and growth timescales are very large and the disk will have disappeared before a protoplanet can gain significant mass~\citep{Lissauer1987,Lissauer1993,Kokubo2002,Mordasini2009a}.
Exactly at what orbital separations efficient planet growth is possible further depends on the amount, size, mass, and aerodynamic properties of planetesimals available there, and thus on the solid disk mass $M_\mathrm{solid,0}$ (see below for a more detailed discussion on the interplay between orbital distance and local planetesimal density).
As can be seen in the lower left panel of Fig. \ref{fig:initial-mass_per_cluster}, intermediate orbits provide the best conditions for rapid growth.
	These trends are responsible for the clear separation of planet clusters in the $a_{\mathrm{start}}$-$M_\mathrm{P}$ plane.
 Very small or very large initial orbital separations always lead to ``failed cores" (low-mass instances of clusters 2 and 4).
    Short-period terrestrial planets and super-Earths (cluster 4) start on small orbits less than \SI{1}{\au}.
	(sub-)Neptunes (cluster 1) require intermediate orbits of roughly \SIrange{0.5}{10}{\au}.
	Finally, giant planets (cluster 3) start on distant orbits ($\gtrsim \SI{3}{\au}$).

Other initial parameters show rather diverse importances that depend on the planet type.
The mean decision boundaries (Eqn \ref{eq:meanDecisionBoundary}) of $M_\mathrm{solid,0}$ and $M_\mathrm{gas,0}$ are close to zero for all clusters except giant planets, implying a small feature importance of these parameters for most planet types.
While these two parameters are correlated in our model, which could in principle spuriously decrease their $MDI$, their relation to $M_\mathrm{P}$ (lower panels of Fig.~\ref{fig:initial-mass_per_cluster}) reveals indeed only a weak relation to planet type.
The picture differs for giant planets, which only form in disks that are rich both in gas ($M_\mathrm{gas,0} \gtrsim \SI{0.04}{\mSun}$) and solids ($M_\mathrm{solid,0} \gtrsim \SI{200}{\mEarth}$).
Given a specific starting location of its core, the efficiency of giant planet formation is strongly governed by $M_\mathrm{solid, 0}$.
The reason is this parameter's direct relation to the local planetesimal density in the disk and thus a protoplanet's ability to reach a core mass sufficient for runaway gas accretion.
Lastly, the disk lifetime stipulates the time within which planet formation has to conclude.
Surprisingly, this parameter shows close to no correlation with the resulting planet type.
This shows that most disks provide material long enough (median~$ \approx\SI{3.4}{\mega\year}$) to complete planet formation.
Within the scope of our model, early disk dispersal is not the preferred pathway to halt planet formation at low and intermediate masses.
	
We conclude that the occurrence of a certain type of planet is fundamentally related to disk properties, and it depends in particular on the orbital distance where the planetary embryo forms.
	Currently, we treat this important parameter as a Monte Carlo variable that is distributed based on simple theoretical arguments~\citep{Kokubo2000}.
	This is a major shortcoming of our formation model and our findings highlight the importance of a consistent treatment of planetary embryo formation~\citep{Voelkel2020b,Voelkel2020c}.
 Another effect we neglected thus far are the gravitational interactions between planets.
	We address this aspect below by discussing simulations done with the same model but multiple forming planets per disk~(see Sect.~\ref{sec:discuss_nbody}).
Future studies should also take into account the effects of pebble accretion~\citep{Ormel2010, Lambrechts2012}, which influence the efficiency of solid accretion and may lead to a global redistribution of solid material in protoplanetary disks~\citep[e.g.,][]{Lambrechts2014a,Morbidelli2015,Ormel2017,Bitsch2019a}.

\subsection{Disk mass and embryo distance as predictors for planet type}
Now that we have identified the solid disk mass and the initial orbital separation of a planetary embryo as the most important features, we investigate the regions different planet types occupy in the space that these parameters span.
Figure~\ref{fig:InitialsPairplot} shows distinct borders between the different clusters that can be explained by the combination of processes our planet formation model covers.
The diagonal border between cluster~1 planets, which correspond to icy and atmosphere-bearing ``Neptunes'' on close and intermediate orbits, and cluster~4 planets, which are dry (super-)Earths, is shaped by photoevaporation of planetary envelopes:
we recall that the clustering algorithm made the separation between these clusters mainly in $R_\mathrm{P}$, which leads to a completely atmosphere-less (super-)Earth cluster and a cluster of Neptunes that predominantly bear H/He envelopes.
However, close to all (super-)Earths initially held an envelope that they subsequently lost due to photoevaporation, a fate that the more massive Neptunes were spared.
Thus, the more solid material is available at a specific orbital distance, the more likely planets will grow massive enough to retain their atmospheres in the long term.
The efficiency of photoevaporation is further a function of orbital distance, leading to the negative slope of the border between clusters~1 and 4 in $a_{\mathrm{start}} - M_\mathrm{solid,0}$~\citep{Jin2018}.
Cluster~2 (``icy cores'') contains only terrestrial planets and failed cores with high amounts of volatile species and no atmospheres.
They formed on distant orbits where the growth timescale is large, preventing them from growing beyond terrestrial size within the lifetime of the protoplanetary disk~\citep{Kokubo2000}.

\subsection{Oligarchic growth of giant planets}\label{sec:oligarchic_giants}
\begin{figure}
		\centering
		\includegraphics[width=\hsize]{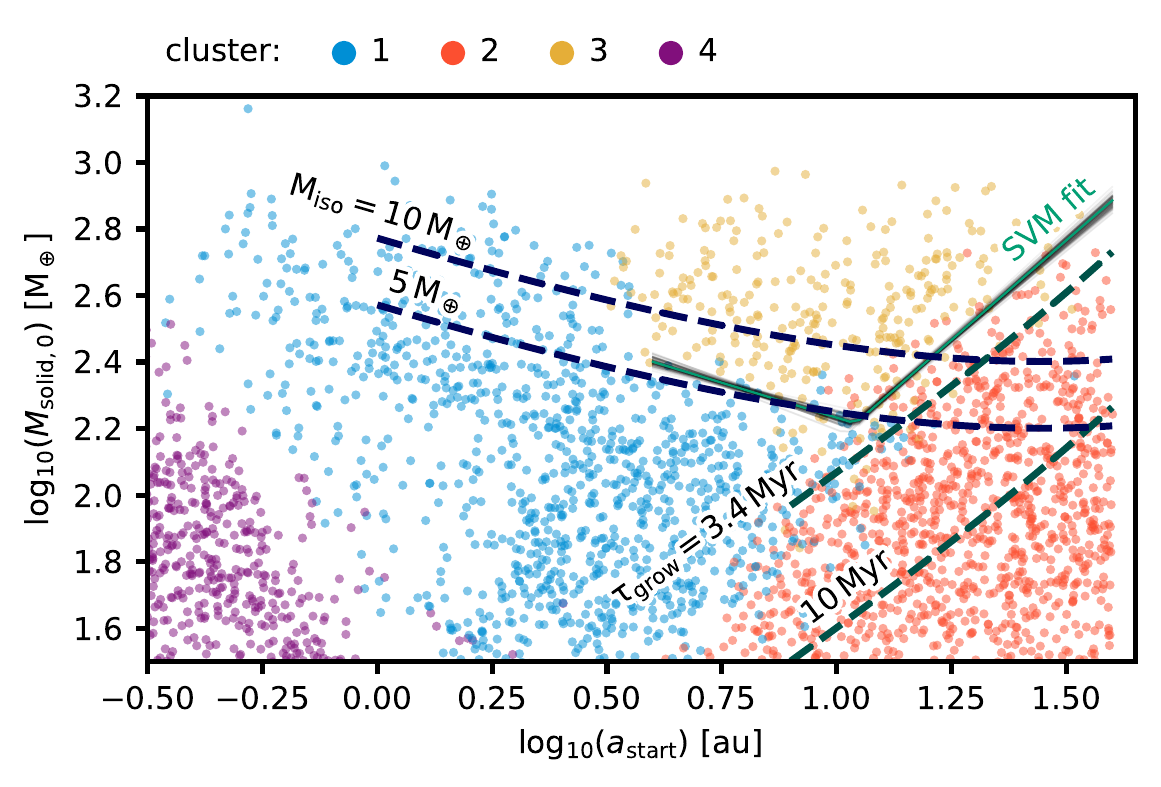}
		\caption{The four clusters of planets in $a_{\mathrm{start}} - M_\mathrm{solid,0}$ space of their nascent protoplanetary disk.
		The green line is the hyperplane that best separates the giant planet cluster (yellow markers) from the other planets and was obtained by training a Support Vector Machine (SVM).
		Closeby gray lines show random draws from bootstrap sampling and illustrate the uncertainties.
		We overplot isolines of planetesimal masses needed to reach specific core masses (blue dashed lines), as well as isolines corresponding to specific growth timescales for reaching a core mass of \SI{10}{\mEarth} (green dashed lines).
		Their slopes are similar to the SVM fit that encloses the giant planets,
		indicating that the onset of runaway growth is limited by the locally available planetesimal mass and by the disk lifetime.}
		\label{fig:Msolid-aStart}
	\end{figure}

	The giant planets (cluster~3) in our planet population occupy a distinct region at large starting positions and high solid disk masses~(see Fig.~\ref{fig:InitialsPairplot}).
	It abruptly cuts off around \SI{4}{\au}, which corresponds to typical water ice line positions at accretion time~\citep{Burn2019}.
	Here, the solid surface density jumps by a factor of four~\citep{Mordasini2012d}, and significantly higher total solid disk masses are required to reach runaway gas accretion interior of this orbit.
	We therefore only considered planets beyond \SI{4}{\au} when we characterized the shape of the giant planet cluster.
We did so by determining the hyperplanes in $a_{\mathrm{start}} - M_\mathrm{solid,0}$ space that best separate these planets from other species.
A Support Vector Machine~\citep[SVM, ][]{Cortes1995} maximizes the distance of this plane to planets that belong to the ``giant planets'' cluster and all those that do not.
We used the implementation in \texttt{scikit-learn}~\citep{scikit-learn} with a linear kernel and otherwise default hyperparameters, and trained the SVM on the full population.
As in logarithmic representation the giant planet cluster has a triangular shape, we can approximate its border by a broken power law.
Setting $y = \log_{10}(M_{\textrm{solid}})$ and $x = \log_{10}(a_{\textrm{start}})$, we fitted the piecewise linear function
	\begin{align}\label{eqn:broken_powerlaw}
	y &=
	\begin{cases}
	k_1 x + y_0 - k_1 x_0 & x \leq x_0  \\
	k_2 x + y_0 - k_2 x_0 & x > x_0
	\end{cases}
	\end{align}
	to separation functions found by the SVM\@.
The best-fit values for these parameters are listed in Tab.~\ref{tab:linear_model_params}.
	\begin{table}
		\centering
		\begin{tabular}{cccc}
			$x_0$ & $y_0$ & $k_1$ & $k_2$\\
			\hline
$1.04^{+0.01}_{-0.01}$ & $2.22^{+0.01}_{-0.01}$ & $-0.42^{+0.04}_{-0.05}$ & $1.20^{+0.03}_{-0.04}$
		\end{tabular}
		\caption{Best-fit parameters for the broken power-law in Equation \ref{eqn:broken_powerlaw}. Uncertainties are 16th and 84th percentiles obtained via bootstrap sampling.}
		\label{tab:linear_model_params}
	\end{table}
We calculated their uncertainties by the bootstrapping method: we repeatedly drew $N$ random planets with replacement, where $N$ is the total number of planets in our synthetic planet population, and trained the SVM on each of 1000 samples generated this way.
In Fig.~\ref{fig:Msolid-aStart}, we overlay the so found giant planet boundary onto the planets in $a_{\mathrm{start}} - M_\mathrm{solid,0}$ space.
Generally, giant planets form when $\log_{10}\left(\frac{M_{\textrm{solid}}}{\SI{1}{\mEarth}}\right) \gtrsim 2.7 - 0.4 \log_{10}\left(\frac{a_{\textrm{start}}}{\SI{1}{\au}}\right)$ for cores emerging within $\sim \SI{10}{\au}$ and when $\log_{10}\left(\frac{M_{\textrm{solid}}}{\SI{1}{\mEarth}}\right) \gtrsim 1.0 + 1.2 \log_{10}\left(\frac{a_{\textrm{start}}}{\SI{1}{\au}}\right)$ for cores emerging beyond.
We point out that this result is only valid in the context of the assumptions of our model.
Plausible limitations that might have influenced this outcome are the assumptions of a single population of planetesimals of the same size and efficient embryo formation throughout the disk, the non-consideration of pebble accretion~\citep{Ormel2010}, and the largely featureless numerical disk that does not allow for ``planet traps''~\citep{Chambers2009}.
Another probable source of error is the omission of gravitational interactions between planets in the same system -- the giant planet domain shifts moderately and is more diffuse when multiple concurrently forming planets are assumed (see Sect.~\ref{sec:discuss_nbody}).
Nevertheless, we focus here on typical outcomes of isolated protoplanets since it allows a more quantitative assessment.

We also compared this boundary to characteristic parameters for planetesimal accretion in the oligarchic growth regime: the planetesimal isolation mass $M_\mathrm{iso}$ and the growth timescale $\tau_\mathrm{grow}$~\citep[e.g.,][]{Kokubo2000,Raymond2014a}.
	On intermediate orbits of a few au, planetary growth is limited by the amount of material that can be accreted.
	$M_\mathrm{iso}$ is a useful concept to quantify the maximum attainable core mass given this limit.
	On the other hand, $\tau_\mathrm{grow}$ gives an estimate for the time needed to reach a certain core mass, and sets the limit for wider orbits.
	For comparison with the giant planet cluster, we computed the local planetesimal densities corresponding to specific values of $M_\mathrm{iso}$ and $\tau_\mathrm{grow}$ and translated them into total planetesimal disk masses $M_\mathrm{solid,0}$.
	See Appendix~\ref{sec:MisoTgrow} for derivations of these quantities.
	
	Since our model includes planet migration, planets can accrete solid material beyond their planetesimal isolation mass by moving through the disk.
	Nevertheless, $M_{\mathrm{iso}}$ is a proxy for how much can be accreted at a specific orbital distance and it is instructive to compare the shape of the giant planet population in $a_{\mathrm{start}} - M_\mathrm{solid,0}$ space with the borders between planet clusters.
	In Fig.~\ref{fig:Msolid-aStart}, we overplot isolines of disk solid masses necessary to reach different planetesimal isolation masses as a function of orbital separation (dashed blue lines).
	The lower border of the giant planet cluster matches well the slope of these lines. %
	This indicates that in intermediate-mass disks with a few hundreds of \SI{}{\mEarth} in solids, giant planet formation is limited by the protoplanets reaching $M_{\mathrm{iso}}$, that is, by clearing their feeding zone from solid material.
	We caution that the proximity of this border to the $M_{\mathrm{iso}}=\SI{5}{\mEarth}$ isoline does not imply that runaway gas accretion has set in at this mass, as planet migration results in a larger effective feeding zone~\citep{Alibert2005}.
	
	Beyond $\sim \SI{10}{\au}$, the border of the cluster matches the slope of isolines for different growth timescales.
	At these larger orbital distances, $\tau_\mathrm{grow}$ can reach the order of \SI{}{\mega\year} for low planetesimal surface densities and thus becomes comparable to the lifetime of the protoplanetary disk.
	In this regime, the growth of a planetary core is limited by the time available to accrete the planetesimals in the domain of a planet's orbit.
	As can be seen in the plot, the $M_\mathrm{solid,0} (a)$ isoline where the growth timescale corresponds to the median of the disk lifetime, $\tau_\mathrm{grow} \approx \SI{3.4}{\mega\year}$, is a good fit to the border between giant planets (yellow) and icy cores (red).
	Indeed, most of the giant planets close to this threshold formed in long-lived disks (see Fig.~\ref{fig:Msolid-aStart_giants_tdisk}).
	This indicates that for planetesimal densities just sufficient for the formation of massive cores, entering runaway gas accretion depends on the longevity of the host disk.

\subsection{The influence of N-body interactions}\label{sec:discuss_nbody}
	Our cluster analysis and prediction from initial conditions has shown that even in the case of multi-planet systems with gravitational interactions, most of the links between disk and planet properties remain intact (see Sect.~\ref{sec:predict_NG76_results}).
    Still, the demographic structures in the multi-planet population are somewhat smeared out compared to the single-planet case, and the strength of this effect is different for individual clusters.
We have seen that (super-)Earths and Neptunes are affected the most by this sort of mixing.
	These planet types cannot be reliably predicted from disk properties if N-body interactions are taken into account.
	Interestingly, the confusion is asymmetric: planets predicted as Neptunes often become (super-)Earths, while those predicted as (super-)Earths rarely become Neptunes.
	The reason is something the classifier cannot predict: the misclassified \SE\ are typically planets that got stripped of their atmospheres in giant collisions with other planets.
	From this follows that our model would produce too many Neptunes if such collisions are not taken into account (as is the case in single-planet simulations).
This highlights the need for global planet formation models to include a consistent treatment of N-body interactions and giant impacts, as has already been suggested by~\citet{Alibert2013} and in \paperone.
	
	Another difference compared to the single-planet case is that close-in planets with small radii and masses are strongly depleted.
	This is because they often undergo giant collisions and merge into more massive bodies.
    The resulting lack of ``sub-Earths'' provides an interesting prediction for future planet searches that will push beyond the current mass/radius limits.
	Whether a multitude or a desert of such planets will be found could give valuable clues to the prevalence of planetary collisions.

\section{Conclusions}\label{sec:conclusion}
We have investigated how different properties of protoplanetary disks relate to the emergence of different planet types in a planetesimal-based core accretion context.
By performing a cluster analysis on synthetic planet populations from a global model of planet formation and evolution, we identified clusters of planets in a parameter space of typical exoplanet observables.
We examined how well these clusters can be predicted from disk properties and studied the dependencies of different planet types.
Our main conclusions are:
\begin{enumerate}
	\item Planets form distinct groups in $\{a, M_\mathrm{P}, R_\mathrm{P}\}$ space, especially when dynamical interactions within multi-planet systems are neglected.
    Without presupposing planet types or their number, we identified four clusters corresponding to \mbox{(sub-)Neptunes}, icy cores, giant planets, and (super-)Earths.
	\item These groups differentiate within the first \SI{0.1}{\mega\year} of the formation process and show correlations with properties of their host disks. Such associations between disk and planet properties enable the prediction of planet species to high accuracy ($98\,\%$ in the single-planet case and $89\,\%$ in the multi-planet case).
	\item The most important predictor for planet clusters is the orbital position of the emerging planetary core, followed by the solid mass available in the disk. The disk lifetime plays a subordinate role, but can be a limiting factor for threshold values of the above mentioned parameters.
	\item The position of giant planets in disk parameter space can be associated with known characteristics of oligarchic planetesimal accretion: for limited available amounts of solid material and within~\SI{\sim10}{\au}, core growth is limited by planetesimal isolation and giant planets form when $\log_{10}\left(\frac{M_{\textrm{solid}}}{\SI{1}{\mEarth}}\right) \gtrsim 2.7 - 0.4 \log_{10}\left(\frac{a_{\textrm{start}}}{\SI{1}{\au}}\right)$.
	On more distant orbits, core accretion is limited by the growth timescale and giants emerge when $\log_{10}\left(\frac{M_{\textrm{solid}}}{\SI{1}{\mEarth}}\right) \gtrsim 1.0 + 1.2 \log_{10}\left(\frac{a_{\textrm{start}}}{\SI{1}{\au}}\right)$.
	\item When multiple planets form and interact in the same system, for most planet types the associations between disk properties and planet properties remain.
	However, planets on track to become sub-Neptunes often lose their atmospheres in giant collisions and turn into super-Earths, which impedes predictions for this planet type.
\end{enumerate}
Overall, we have shown that synthetic planet populations from state-of-the-art core accretion models largely mirror the planet types recognized by exoplanet demographics.
Our results highlight the importance of N-body integrations in global planet formation models that aim for reliable predictions in the domain of low-mass planets. 
Beyond that, constraining the orbital distances at which planetary cores form is of major relevance for the full range of planet types.
Population syntheses of the next generation should recognize this by including self-consistent treatments of planetary embryo formation.

\begin{acknowledgements}
The authors thank Gabriele Pichierri for fruitful discussions.
We thank the anonymous referee for valuable comments that improved the manuscript.
This work was supported by the DFG Research Unit FOR2544 “Blue Planets around Red Stars”, project no. RE 2694/4-1.
T.H. acknowledges support from the European Research Council under the Horizon 2020 Framework Program via the ERC Advanced Grant Origins 83 24 28.
This research was supported by the Deutsche Forschungsgemeinschaft through the Major Research Instrumentation Programme and Research Unit FOR2544 “Blue Planets around Red Stars” for T.H. under contract DFG He 1935/27-1 and for H.K. under contract DFG KL1469/15-1.
Parts  of  this  work  has  been  carried  out  within  the  framework of  the  National  Centre  for  Competence  in  Research  PlanetS  funded  by  the Swiss  National  Science  Foundation  (SNSF).
R.B.  and  Y.A.  acknowledge financial support from the SNSF under grant $200020\_ 172746$.
Some of the computations have been carried out on the DRACO cluster of the Max Planck Society, which is hosted at the Max Planck Computing and Data Facility in Garching (Germany).
\end{acknowledgements}

\bibliographystyle{aa} %
\bibliography{PhD,addLit,unsupervised} %

\begin{appendix}

\section{The choice of a clustering algorithm}\label{sec:appendix_cluster_analysis}
\subsection{Clustering algorithms}\label{sec:appendix_clusteralgs}
\rev{For the cluster analysis in Sect.~\ref{sec:cluster_analysis}, we examined several other clustering algorithms in addition to GMM\footnote{\texttt{sklearn.mixture.GaussianMixture}} and explored their behavior on our data~set.
For each method, we used its implementation in \texttt{scikit-learn}~\citep{scikit-learn} and, where applicable, chose the default Euclidean distance metric.
The algorithms considered are centroid, density, or hierarchical-based.
A centroid-based method we explored was K-means~\citep{MacQueen1967, Lloyd1982}.
In the density-based group, we tested DBSCAN and OPTICS~\citep{Ester1996, Ankerst1999}.
For hierarchical clustering, we examined Agglomerative clustering~\citep{Ward1963} besides GMM~\citep{McLachlan1988}.
}

K-Means\footnote{\texttt{sklearn.cluster.KMeans}}\citep{MacQueen1967, Lloyd1982} is a centroid-based clustering algorithm: it randomly initializes $k$ centroids and associates each data point to the centroid that is closest to it, then shifts the centroids to the mean of their cluster.
These steps are repeated until no changes occur.
The algorithm requires only a single hyperparameter $k$, which is the number of clusters.

Agglomerative clustering\footnote{\texttt{sklearn.clustering.AgglomerativeClustering}} \citep{Ward1963} is a bottom-up hierarchical clustering algorithm: each data point begins as its own cluster and incrementally merges similar pairs of clusters into a new cluster. This process is repeated until there are $k$ clusters left, where $k$ is the hyperparameter for the number of clusters.
When testing this algorithm, we used a hyperparameter called linkage to quantify `similarity' between pairs of clusters \citep[e.g.,][]{Ward1963, Szekely2005}.
Empirically, we found that the \rev{``Ward''} linkage is optimal.

DBSCAN\footnote{\texttt{sklearn.cluster.DBSCAN}}~\citep{Ester1996} is a density-based clustering algorithm classifying each data point as either a core point (with at least \texttt{minPts} neighboring points within a distance $\epsilon$), a reachable point (that is within distance $\epsilon$ of the core point), or an outlier (that is not reachable by any core point).
All core points and their reachable points form a cluster, but outliers do not.
\rev{The method we tested is an advancement of DBSCAN with improved performance on data sets of varying density.
This method called OPTICS\footnote{\texttt{sklearn.cluster.OPTICS}}~\citep{Ankerst1999}} has one hyperparameter: \texttt{minPts} -- the minimum number of points nearby to make a core point.

\subsection{Validation metrics \rev{and choice of method}}\label{sec:appendix_metrics}
\rev{Each of these methods has hyperparameters, that is, parameters that are not derived during model training but that control the learning process itself.
We used a number of validation metrics to quantify the clustering performance for each method and specific choice of hyperparameters.
}
Some of these metrics are method-specific and can only be used with a specific algorithm.
These are the elbow method~\citep[e.g.,][]{Thorndike1953, Ketchen1996}, the Bayesian and Aikake Information Criterions~\citep[BIC and AIC, e.g.,][]{Akaike1973, Schwarz1978, Cavanaugh2019}, and the dendrogram method~\citep[e.g.,][]{Nielsen2016}.
The elbow method is used to evaluate the performance of the K-Means algorithm.
By plotting the within-cluster sum-of-squares against $k$, an `elbow'-shaped curve emerges.
The ideal $k$ will be one close to the `elbow'.
The reasoning for this is that we aim to find the first $k$ that minimizes the within-cluster sum-of-squares.
BIC and AIC are used for GMM. Both are based on information theory and are used to prevent overfitting and underfitting to choose the most optimized model.
The dendrogram method is used to judge the bottom-up process of Agglomerative clustering.
It shows the clustering at each hierarchy, where the y-axis is the distance between clusters and the x-axis shows the clusters.
Therefore, the goal is to perform a horizontal cut such that the vertical distance is maximized.
\rev{As one traverses} up the hierarchy, the vertical distance naturally increases.

In addition to these scores, \rev{we used} the following scalar-valued metrics that can be used for any method:
the Silhouette score~\citep{Rousseeuw1987}, the Caliński-Harabasz score~\citep[CH, ][]{Calinski1974}, and the Davies-Bouldin score~\citep[DB, ][]{Davies1979}.
The Silhouette score is computed from the mean intra-cluster distance and the mean nearest-cluster distance.
Silhouette scores range between -1 and 1 with 1 being the best and -1 being the worst, and values near 0 implying overlapping clusters.
We aimed to maximize this score.
The Caliński-Harabasz score is the ratio of the within-cluster dispersion and the between-cluster dispersion, where dispersion is the sum of the squared distances.
Again, we aimed to maximize this score.
The Davies-Bouldin score determines the clustering performance by using the ratio of the within-cluster distances to the between-cluster distances.
As a result, compact clusters that are far apart give better scores.
The minimum score is 0, and we aimed to minimize this score.

\rev{
\subsection{Model selection}
Our approach in selecting the best clustering method was as follows:
first, we applied each method to the $\{a, M_\mathrm{P}, R_\mathrm{P}\}$ subspace of the \rev{\textit{NG73} planet population} for a wide range of hyperparameters.
We then compared the validation metrics computed for the resulting clusterings.
The scores did not always agree unanimously, which is expected, as the structures in our multidimensional data~set are rather complex and the scores consider different goals \rev{regarding} an optimal clustering.
The next step was thus to produce, for each combination of method and hyperparameters, scatter plots that showed the clustering results in different projections of $\{a, M_\mathrm{P}, R_\mathrm{P}\}$ space.
Using these plots, we could compare the different partitionings and determine the most sensible model.
}
\begin{figure*}
	\centering
	\includegraphics[width=\hsize]{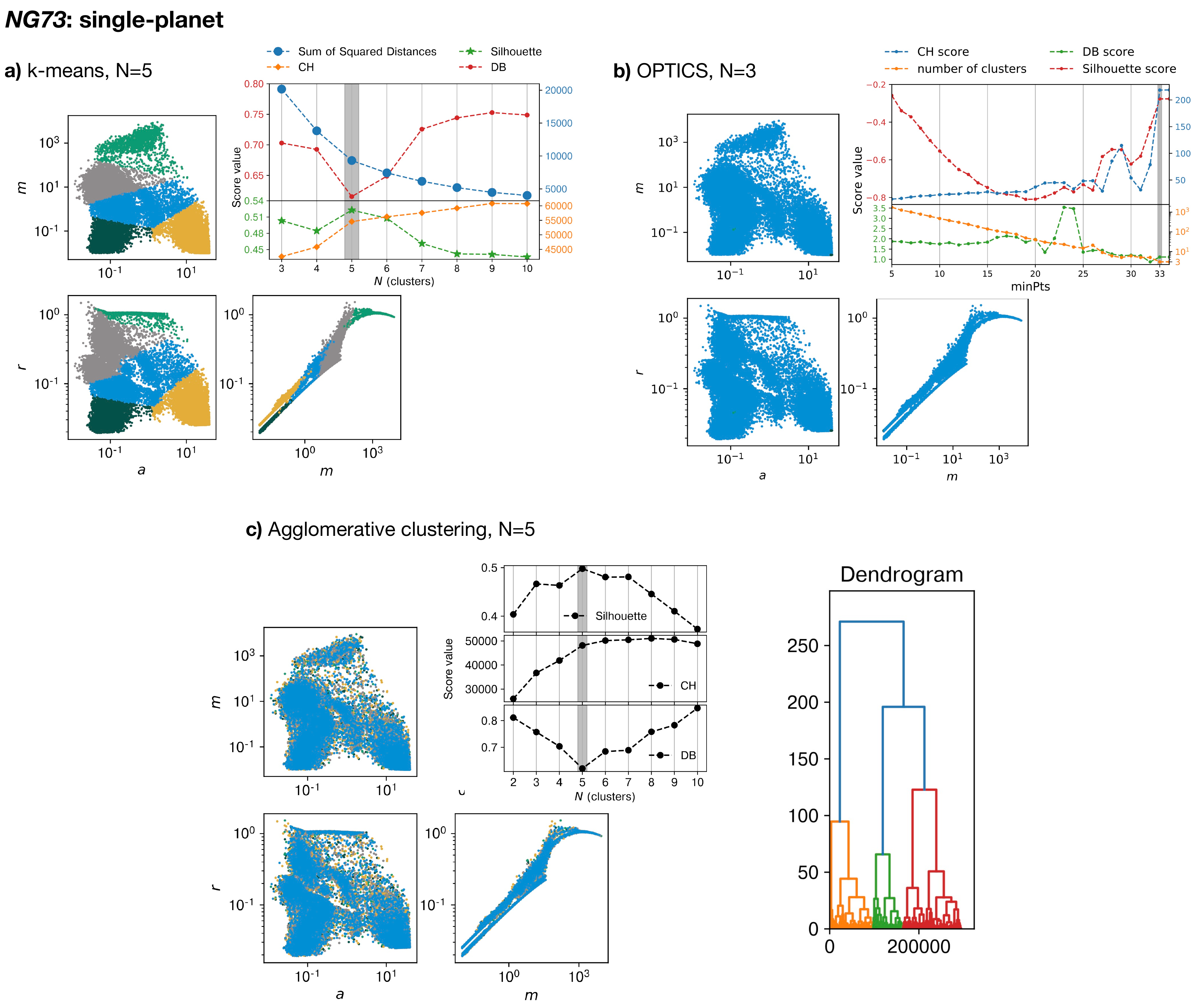}
	\caption{\rev{Diagnostic plots for clustering method selection.
	For each alternative clustering algorithm we explored, we show the validation metrics we used to choose hyperparameters.
	Based on these metrics, we show the resulting clustering for the most promising choices in the corner plots.
a) Even in the best case ($N=5$), k-means' approach to draw cluster borders is too simplistic to account for the structure in our data.
	b) For the numerically best choice of $minPts$, OPTICS finds three clusters of extremely different sizes. Most of the data belong to a single cluster that covers the whole domain, and no sensible relation to the data point density is apparent.
	c) Agglomerative clustering suggests the existence of five clusters. Again, no reasonable partitioning is visible. The lower right panel shows the dendrogram corresponding to this clustering.
	}}
	\label{fig:diagnostics_non-GMM}
\end{figure*}
\rev{Figure~\ref{fig:diagnostics_non-GMM} shows these diagnostic plots for k-means, OPTICS, and Agglomerative clustering, using the choice of hyperparameters considered most appropriate.
The diagnostic plots for GMM are shown in Fig.~\ref{fig:diagnostics_GMM}.
Based on this selection procedure, GMM showed the best performance and we considered it our nominal method for clustering.
}

\rev{
A free parameter of GMMs is the number of components $N$, which we chose using the same two-step approach as in the method selection.
After the validation metrics suggested $N=4, N=6$ for \textit{NG73} and $N=3, N=5$ for \textit{NG76}~(see Fig.~\ref{fig:GMM_scores}), we assessed the diagnostic plots shown in Fig.~\ref{fig:diagnostics_GMM}.
\begin{figure*}
	\centering
	\includegraphics[width=\hsize]{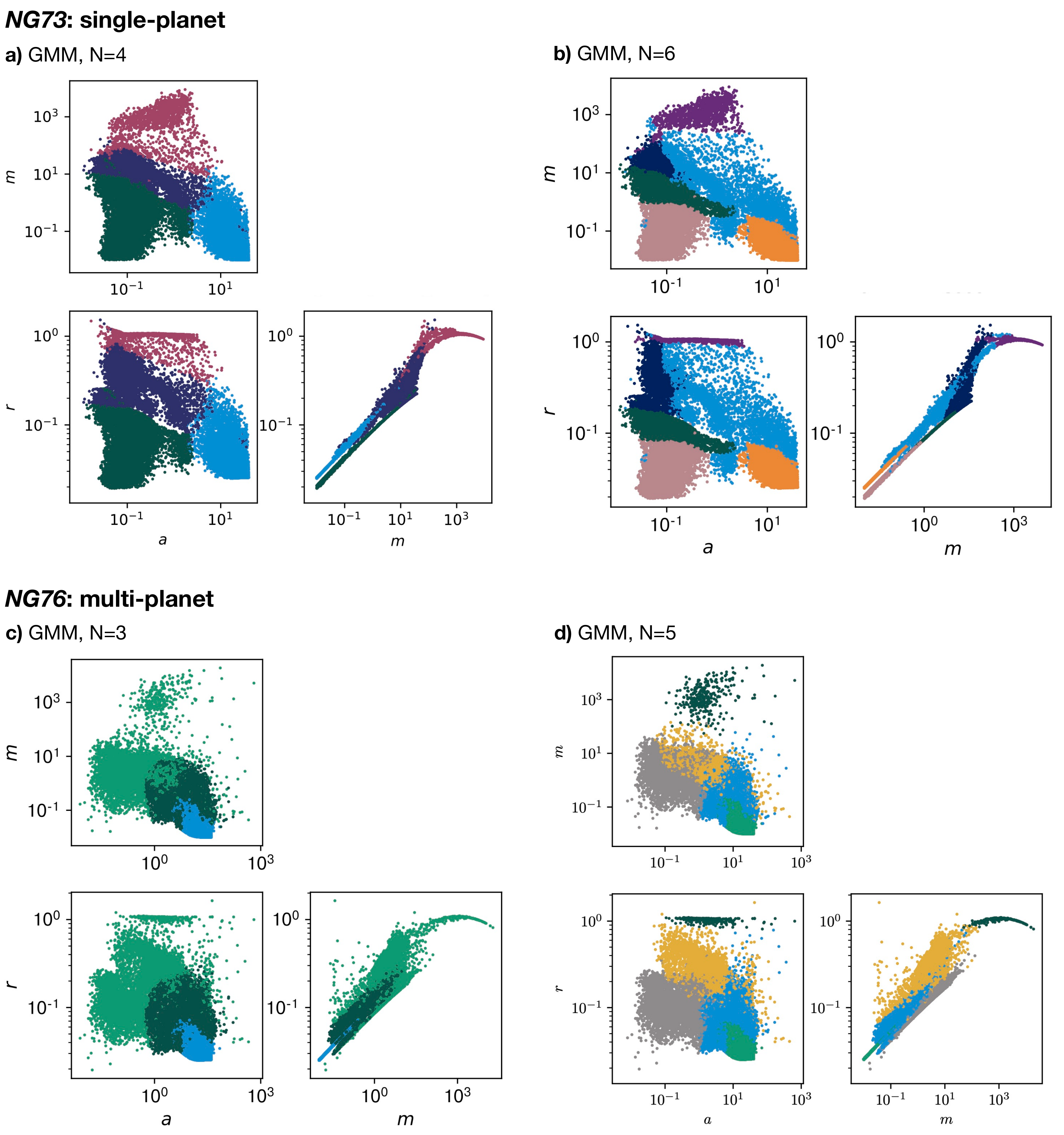}
	\caption{\rev{Diagnostic plots for GMM clustering model selection.
   According to our validation metrics, the best candidate number of clusters are $N=4, N=6$ for \textit{NG73} and $N=3, N=5$ for \textit{NG76}~(compare Fig.~\ref{fig:GMM_scores}). The panels a)--d) show the clustering results of these choices.
The models in a)~($N=4$) and d)~($N=5$) trace the over- and underdensities in the domain best and we consider them our nominal models.}
}
	\label{fig:diagnostics_GMM}
\end{figure*}
For \textit{NG73}, we found that the GMM with $N=6$ reaches similar scores than $N=4$ but traces less reliably the underdensities in the domain and partly draws cluster borders through rather arbitrary regions.
We thus chose the GMM with $N=4$ as our nominal model for the single-planet case.
For \textit{NG76}, the model with more components reliably detects visible overdensities and outperforms the less complex model.
Hence, we adopted the GMM with $N=5$ as the nominal model for the multi-planet case.
}

\section{Boundary conditions for giant planet formation}\label{sec:MisoTgrow}
\subsection{Derivation of isolation mass and growth timescale}
In Sect.~\ref{sec:oligarchic_giants}, we characterize the cluster of giant planets in $a_{\mathrm{start}} - M_\mathrm{solid,0}$ space, where it occupies a distinct triangular region.
In the following, we derive two quantities that shape this region: the total solid disk mass as a function of orbital distance for different planetesimal isolation masses and for different growth timescales.

$M_\mathrm{iso}$ gives the mass in planetesimals a protoplanet can accrete given a feeding zone of width $b \simeq 10 r_H$, where $r_H = a \left( \frac{M_\mathrm{P}}{3 M_\star}\right)^{1/3}$.
Then,
\begin{align}
	M_\mathrm{iso} &= 2\pi a b \Sigma_\mathrm{solid} \simeq 2\pi a 10 a\left( \frac{M_\mathrm{P}}{3 M_\star}\right)^{1/3}  \Sigma_\mathrm{solid},\\
	\intertext{where $\Sigma_\mathrm{solid}$ is the planetesimal surface density.
	Setting the planetary mass to the planetesimal isolation mass, $M_\mathrm{P} \equiv M_\mathrm{iso}$, yields}
	M_\mathrm{iso} &= \left(\frac{20 \pi }{3^{1/3}}\right)^{3/2} a^3 \Sigma_\mathrm{solid}^{3/2} M_\star^{-1/2}.
\end{align}
To get an estimate on which initial solid mass content is required to reach a certain isolation mass, we express this as
\begin{equation}
	\label{eq:Sigsolid}
	\Sigma_\mathrm{solid} = \left(\frac{3^{1/3}}{20\pi}\right) \frac{M_\star^{1/3} M_\mathrm{iso}^{2/3}}{a^2}.
\end{equation}
For the power law disk profile used in our model~\citep{Andrews2009},
\begin{equation}
	\Sigma(r) = \Sigma_0 \left(\frac{r}{r_0}\right)^{-\beta} \exp \left[- \left(\frac{r}{r_\mathrm{cut,g}}\right)^{(2-\beta)}\right],
\end{equation}
we consider the outer disk radii $r_\mathrm{cut,g}$ and $r_\mathrm{cut,s}$ for the gas and solid disk, respectively.
The radial slope of $\Sigma_\mathrm{solid}$ is characterized by the power law index $\beta$, and $\Sigma_0$ is the surface density at a reference orbital distance $r_0 = \SI{5.2}{\au}$.
Then, the total mass of the planetesimal disk is
\begin{equation}
	\label{eq:Msolid}
	M_\mathrm{solid} = \frac{2\pi\Sigma_0}{r_0^{-\beta}} \frac{r_\mathrm{cut,s}^{2-\beta}}{2-\beta}\,,
\end{equation}
where $r_\mathrm{cut,s} = 0.5 r_\mathrm{cut,g}$~\citep[following findings from dust disk observations, ][]{Ansdell2018} and $\beta=1.5$ \citep[motivated by planetesimal formation models, ][]{Lenz2019}.
Substituting Equation~\ref{eq:Sigsolid} into Equation~\ref{eq:Msolid}, the total solid mass required to reach $M_{\mathrm{iso}}$ is given by
\begin{equation}
	M_\mathrm{solid} (M_\mathrm{P} = M_\mathrm{iso}) = \frac{3^{1/3}}{10} \frac{r_\mathrm{cut,s}^{2-\beta}}{2-\beta} \frac{M_\star^{1/3} M_\mathrm{iso}^{2/3}}{a^{2-\beta}} \exp \left[ -\left(\frac{a}{r_\mathrm{cut,s}}\right)^{2 - \beta} \right]^{-1}
\end{equation}

Similarly, we can derive the solid disk mass needed to reach a specific mass in the outer disk regions, where growth is mainly limited by the growth timescale $\tau_{\mathrm{grow}}$.
For the oligarchic growth regime~\citep{Ida1993}, this timescale can be approximated by
\begin{equation}\label{eq:taugrow}
\begin{split}
	\tau_\mathrm{grow} \approx 1.2 \times 10^{5} \mathrm{yr}\left(\frac{\Sigma_{\mathrm{p}}}{10 \mathrm{~g} \mathrm{~cm}^{-2}}\right)^{-1}\left(\frac{a}{1 \mathrm{au}}\right)^{1 / 2}\left(\frac{M_{c}}{M_{\oplus}}\right)^{1 / 3}\left(\frac{M_{\star}}{M_{\odot}}\right)^{-1 / 6} \\
	\times \left[\left(\frac{\Sigma_{\mathrm{g}}}{2400 \mathrm{~g} \mathrm{~cm}^{-2}}\right)^{-1 / 5}\left(\frac{a}{1 \mathrm{au}}\right)^{1 / 20}\left(\frac{M_\mathrm{pla}}{10^{18} \mathrm{~g}}\right)^{1 / 15}\right]^{2}
\end{split}
\end{equation}
~\citep{Mordasini2018}.
Solving for $\Sigma_{\mathrm{p}}$ and substituting into Equation~\ref{eq:Msolid} gives
\begin{multline}
	M_\mathrm{solid}(a, \tau_\mathrm{grow}) =
	\SI{7.54}{\gram\per\centi\meter\squared} \frac{r_0^{\beta} r_\mathrm{cut,s}^{2 - \beta}}{2 - \beta}  \left(\frac{M_{c}}{M_{\oplus}}\right)^{1 / 3}  \left(\frac{M_{\star}}{M_{\odot}}\right)^{-1/6}\\
	\times \left[\left(\frac{\Sigma_{\mathrm{g}}(a)}{2400 \mathrm{~g} \mathrm{~cm}^{-2}}\right)^{-1 / 5}\left(\frac{M_\mathrm{pla}}{10^{18} \mathrm{~g}}\right)^{1 / 15}\right]^{2} \left(\frac{\tau_\mathrm{grow}}{\SI{1}{\mega\year}} \right)^{-1} \left(\frac{a}{1 \mathrm{au}}\right)^{3/5},
\end{multline}
where $\Sigma_{\mathrm{g}}$ was computed using the population-wide median of the reference surface density $\Sigma_\mathrm{0,gas}$.
For the cutoff radii of the gas and solid disk, we proceeded in the same way and assumed the population median, respectively.
For the planetesimal mass $M_\mathrm{pla}$, we assumed a density of \SI{1}{\gram\per\centi\meter\cubed}, which results in $M_\mathrm{pla} = \SI{1.13e11}{\kilo\gram}$ for the planetesimals in our model~\citep{Emsenhuber2020}.
We adopted a core mass $M_{c}$ of \SI{10}{\mEarth}.

\subsection{Disk lifetime limits giant planet growth}
		\begin{figure}
		\centering
		\includegraphics[width=\hsize]{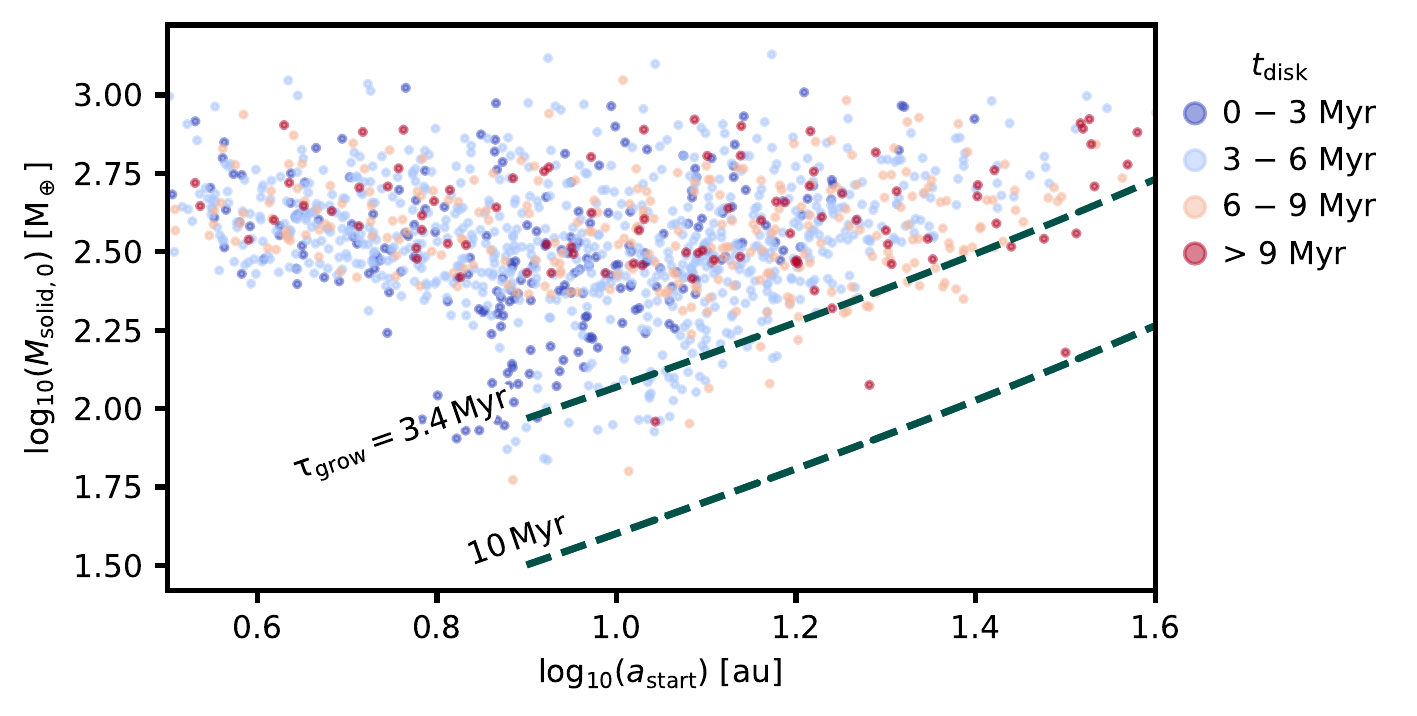}
		\caption{Planetesimal disk mass and initial planet core position of giant planets forming in disks of different lifetimes.
		Markers show the distribution of all planets classified as ``cluster 3: giant planet'' in $a_{\mathrm{start}} - M_\mathrm{solid,0}$ space, color-coded by the lifetime of their nascent disk.
		We overplot isolines of planetesimal masses corresponding to specific growth timescales $\tau_\mathrm{grow}$ for reaching a core mass of \SI{10}{\mEarth}.
        Giant planet growth is limited by the disk lifetime, and the formation of giant planets far out requires high planetesimal masses and long lifetimes.
		}
		\label{fig:Msolid-aStart_giants_tdisk}
	\end{figure}
	
Figure~\ref{fig:Msolid-aStart_giants_tdisk} shows the cluster of giant planets in the space spanned by two important initial disk properties, $a_{\mathrm{start}}$ and  $M_\mathrm{solid,0}$.
The colors correspond to different lifetimes of the protoplanetary disk in which they formed.
Most giants grow (and survive) in disks with lifetimes \SIrange{3}{6}{\mega\year}.
Only long-living disks enable formation of giant planets at low solid disk masses and large orbital distances.
In short-lived disks, there is only a narrow region of embryo starting positions where giant planets grow at low planetesimal surface densities.

\end{appendix}

\end{document}